\documentclass[usenatbib]{mn2e}
\usepackage{graphicx,fixltx2e}

\def\rpd{\hbox{rad\,d$^{-1}$}}

\def\chisq{\hbox{$\chi^2$}}
\def\msun{\hbox{${\rm M}_{\odot}$}}
\def\mspy{\hbox{${\rm M}_{\odot}$\,yr$^{-1}$}}
\def\rsun{\hbox{${\rm R}_{\odot}$}}
\def\rcor{\hbox{$r_{\rm C}$}}
\def\rmag{\hbox{$r_{\rm A}$}}
\def\mstar{\hbox{$M_{\star}$}}
\def\rstar{\hbox{$R_{\star}$}}
\def\teff{\hbox{$T_{\rm eff}$}}
\def\logg{\hbox{$\log g$}}
\def\sn{\hbox{S/N}}
\def\vrad{\hbox{$v_{\rm rad}$}}

\def\eps{\hbox{erg\,s$^{-1}$}}
\def\epspcs{\hbox{erg\,s$^{-1}$\,cm$^{-2}$}}
\def\kms{\hbox{km\,s$^{-1}$}}
\def\vsini{\hbox{$v \sin i$}}

\def\pcc{\hbox{cm$^{-3}$}}

\def\ptt{\hbox{$10^{-4} I_{\rm c}$}}

\def\degr{\hbox{$^\circ$}}

\def\Mdot{\hbox{$\dot{M}$}}
\def\omeq{\hbox{$\Omega_{\rm eq}$}}
\def\dom{\hbox{$d\Omega$}}

\newcommand{\caii}{\hbox{Ca$\;${\sc ii}}}
\newcommand{\fei}{\hbox{Fe$\;${\sc i}}}
\newcommand{\hei}{\hbox{He$\;${\sc i}}}

\begin{document}

\title[Magnetic fields and accretion flows on the cTTS V2129~Oph]{Magnetic fields 
and accretion flows on the classical T~Tauri star V2129~Oph\thanks{Based on observations 
obtained at the Canada-France-Hawaii Telescope (CFHT) which is operated by the National 
Research Council of Canada, the Institut National des Sciences de l'Univers of the Centre 
National de la Recherche Scientifique of France, and the University of Hawaii.} }

\makeatletter

\def\newauthor{%
  \end{author@tabular}\par
  \begin{author@tabular}[t]{@{}l@{}}}
\makeatother
 
\author[J.-F.~Donati et al.]
{\vspace{1.7mm}
J.-F.~Donati$^1$\thanks{E-mail: donati@ast.obs-mip.fr (J-FD); 
mmj@st-andrews.ac.uk (MMJ); 
sg64@st-andrews.ac.uk (SGG); 
petit@ast.obs-mip.fr (PP); 
jerome.bouvier@obs.ujf-grenoble.fr (JB); 
catherine.dougados@obs.ujf-grenoble.fr (CD); 
francois.menard@obs.ujf-grenoble.fr (FM); 
acc4@st-andrews.ac.uk (ACC); 
th@astro.ex.ac.uk (TJH); 
s.v.jeffers@phys.uu.nl (SVJ); 
fpaletou@ast.obs-mip.fr (FP)}, 
M.M.~Jardine$^2$, S.G.~Gregory$^2$, P.~Petit$^1$, J.~Bouvier$^3$, C.~Dougados$^3$, \\ 
\vspace{1.7mm}
{\hspace{-1.5mm}\LARGE\rm 
F.~M\'enard$^3$, A.C.~Cameron$^2$, T.J.~Harries$^4$, S.V.~Jeffers$^{1,5}$ and F. Paletou$^1$ } \\ 
$^1$ LATT, CNRS--UMR 5572, Obs.\ Midi-Pyr\'en\'ees, 14 Av.\ E.~Belin, F--31400 Toulouse, France \\
$^2$ School of Physics and Astronomy, Univ.\ of St~Andrews, St~Andrews, Scotland KY16 9SS, UK \\
$^3$ LAOG, CNRS--UMR 5573, Obs.\ de Grenoble, 31 rue de la Piscine, F--38041 Grenoble, France \\ 
$^4$ School of Physics, Univ.\ of Exeter, Stocker Road, Exeter EX4~4QL, UK \\ 
$^5$ Sterrenkundig Instituut, Univ.\ Utrecht, PO Box 80000, NL--3508 TA Utrecht, The Netherlands
}

\date{2006, MNRAS, submitted}
\maketitle
 
\begin{abstract}  
From observations collected with the ESPaDOnS spectropolarimeter, we report the
discovery of magnetic fields at the surface of the mildly accreting classical 
T~Tauri star (cTTS) V2129~Oph.  Zeeman signatures are detected, both in
photospheric lines and in the emission lines formed at the base of the
accretion funnels linking the disc to the protostar, and monitored over the
whole rotation cycle of V2129~Oph.  We observe that rotational modulation
dominates the temporal variations of both unpolarised and circularly polarised
line profiles.  

We reconstruct the large-scale magnetic topology at the surface of V2129~Oph
from both sets of Zeeman signatures simultaneously.  We find it to be rather
complex, with a dominant octupolar component and a weak dipole of strengths 
1.2 and 0.35~kG respectively, both slightly tilted with respect to the rotation
axis.  The large-scale field is anchored in a pair of 2~kG unipolar radial
field spots located at high latitudes and coinciding with cool dark polar
spots at photospheric level.  This large-scale field geometry is unusually
complex compared to those of non-accreting cool active subgiants with moderate
rotation rates.  

As an illustration, we provide a first attempt at modelling the magnetospheric 
topology and accretion funnels of V2129~Oph using field extrapolation.  We
find that the magnetosphere of V2129~Oph must extend to about 7~\rstar\ to
ensure that the footpoints of accretion funnels coincide with the
high-latitude accretion spots on the stellar surface.   It suggests that the
stellar magnetic field succeeds in coupling to the accretion disc as far out
as the corotation radius, and could possibly explain the slow rotation of
V2129~Oph.  The magnetospheric geometry we derive qualitatively reproduces the
modulation of Balmer lines and produces X-ray coronal fluxes typical of those
observed in cTTSs.  

\end{abstract}


\section{Introduction} 

T~Tauri stars (TTSs) are young low-mass stars that have emerged from their
natal molecular cloud core.   Among them, classical TTSs (cTTSs) are those
still surrounded by accretion discs.  CTTSs are strongly  magnetic;  their
fields are thought to be  responsible for disrupting the central regions of
their accretion discs and  for channelling the disc material towards the
stellar surface along  discrete accretion funnels. This magnetically-channelled
accretion can  determine the angular momentum evolution of protostars 
\citep[e.g.,][]{Konigl91, Shu94, Cameron93, Romanova04} as well as their  
internal structure.  Studying magnetospheric accretion processes of cTTSs 
through both observations and simulations is thus crucial for our 
understanding of stellar formation.  


The Zeeman  broadening of unpolarised spectral lines \citep[e.g.,][]{Johns99a,
Yang05},  has been very successful at determining mean surface magnetic field
strengths on cTTSs, even for complex field topologies, but is rather
insensitive to the large-scale structure of the field.   Spectropolarimetric
techniques, aimed at detecting polarised Zeeman signatures  in spectral lines,
are sensitive to the vector properties of the magnetic field  but  encounter
difficulties in reconstructing highly-complex multipolar field structures.
Until now, circular spectropolarimetry has been used, for cTTSs, to detect and
estimate magnetic fields in accretion funnels of cTTSs \citep{Johns99b,
Valenti04}, but not the photospheric fields \citep[e.g.,][]{Valenti04, Daou06}.  

Here we report a new study of the cTTS V2129~Oph (SR~9, AS~207A,  GY~319,
ROX~29, HBC~264), the brightest TTS in the $\rho$~Oph cloud, using  ESPaDOnS,
the new generation spectropolarimeter recently installed at the  3.6-m
Canada--France--Hawaii Telescope (CFHT).   After briefly reviewing the 
fundamental parameters of V2129~Oph (Sec.~\ref{sec:par}), we present our 
observations (Sec.~\ref{sec:obs}) and describe the shape and rotational
modulation  of the Zeeman signatures detected in both photospheric and emission
lines  (Sec.~\ref{sec:zee}).  Using tomographic techniques, we model the
distribution  of active regions, accretion spots and magnetic fields at the
surface of V2129~Oph  (Secs.~\ref{sec:dim} \& \ref{sec:zdi}).  As an illustration, 
we provide a first attempt at modelling the magnetosphere and accretion funnels of 
V2129~Oph ((Sec.~\ref{sec:mag}).  We finally discuss the implications of our results 
for stellar formation (Sec.~\ref{sec:dis}).

\section{The cTTS V2129~Oph}
\label{sec:par}

Several estimates of the photospheric temperature \teff\ of V2129~Oph are
available  in the literature.  \citet{Bouvier90} and \citet{Bouvier92}
initially suggested that  $\teff\simeq 4200$~K. \citet{Padgett96}
and \citet{Eisner05} proposed  higher temperatures (4650 and 4400~K
respectively) while \citet{Geoffray01} and  \citet{Doppmann03} concluded that
the star is actually cooler (4000 and 3400~K  respectively).  These differences
probably arise because the spectral energy distribution  includes a strong
infrared contribution from the accretion disc \citep{Eisner05} and,  to a lesser
extent, from its companion star\footnote{V2129~Oph is a distant  binary whose
secondary star, presumably a very low-mass star or a brown dwarf  according to
\citet{Geoffray01}, is about 50 times fainter than the  cTTS primary in the V
band.}.  The photometric colours of V2129~Oph  are, moreover, affected by
interstellar reddening, making photometric temperature  estimates slightly
unreliable.  The spectroscopic measurements by \citet{Padgett96} and  
\citet{Bouvier92} appear safest;  we thus choose 4500~K as the most likely  
estimate, with a typical error bar of about 200~K.  

From this temperature estimate, and given the $B-V$, $V-R_{\rm c}$ and
$V-I_{\rm c}$  photometric colours of V2129~Oph (respectively equal to 1.25,
0.80 and 1.60,  \citealt{Bouvier90}), we infer from Kurucz synthetic colours
(including interstellar  reddening, \citealt{Kurucz93}) that the $B-V$ colour
excess $E(B-V)\simeq0.2$ (i.e.\  that the interstellar extinction
$A_V\simeq0.6$) and that the bolometric correction  (including interstellar
extinction) is $-1.3$.   Although V2129~Oph was too faint to be observed with
Hipparcos (and thus to have  its parallax estimated accurately), its membership
of the $\rho$~Oph dark cloud  indicates that its distance is about
$140\pm10$~pc \citep{Bontemps01}.  

\citet{Bouvier90} and \citet{Shevchenko98} find that the maximum  brightness of
V2129~Oph (at minimum spot coverage) is about 11.20;   its unspotted $V$
magnitude is thus $11.0\pm0.2$, given that active stars  usually feature as
much as 20\% spot coverage even at brightness maximum 
\citep[e.g.,][]{Donati03}.  From this, we obtain that the bolometric magnitude 
of V2129~Oph is $4.0\pm0.25$, i.e.\ that its logarithmic luminosity (with
respect  to the Sun) is $0.3\pm0.1$.  Given the assumed \teff, we derive that
V2129~Oph  has a radius of $\rstar=2.4\pm0.3$~\rsun. In Sec.~\ref{sec:dim} we
measure the projected  equatorial rotation velocity to be
$\vsini=14.5\pm0.3$~\kms, where $i$ is the angle between the rotation  axis and
the line of sight. This agrees well with published estimates of
$15.2\pm0.9$~\kms\  and $15.8\pm1.5$~\kms\ by \citealt{Eisner05} and
\citealt{Bouvier90} respectively.  From \vsini\ and the rotation period of
6.53~d \citep{Shevchenko98}, we obtain  that $\rstar \sin
i=1.87\pm0.04$~\rsun.   We infer $i\simeq50\degr$, in good agreement with  the
value of 45\degr\ derived from tomographic modelling of our data (see 
Sec.~\ref{sec:dim}).  

Fitting the evolutionary models of \citet{Siess00} to these parameters, we
infer  that V2129~Oph is a $1.35\pm0.15$~\msun\ star with an age of about
2~Myr and is no longer fully convective, with a small radiative core  of mass
$\simeq0.1$~\mstar\ and radius $\simeq0.2$~\rstar.  

From the equivalent width of the 866~nm \caii\ line emission core (about
0.05~nm, see  Sec.~\ref{sec:zee}, implying a line emission flux of
$10^{-6}$~\epspcs), we obtain  that the mass-accretion rate \Mdot\ of V2129~Oph
is about $4\times10^{-9}$~\mspy\  \citep{Mohanty05}.  \citet{Eisner05} suggests
that \Mdot\ is significantly larger,  about $3\times10^{-8}$~\mspy.  We assume
here that $\Mdot\simeq10^{-8}$~\mspy, i.e.,  a conservative compromise between
both estimates and a typical value for a mildly  accreting cTTS.  

Rotational cycles $E$ are computed according to the ephemeris:  
\begin{equation}
\mbox{HJD} = 2453540.0 + 6.53 E, 
\label{eq:eph}
\end{equation}
where the rotation period is that determined by \citet{Shevchenko98}.  

\begin{table}
\caption[]{Journal of observations.  Columns 1--4 sequentially 
list the UT date, the heliocentric Julian date 
and UT time (both at mid-exposure), and the peak
signal to noise ratio (per 2.6~\kms\ velocity bin) of each
observation.  Column 5 lists the rms noise level (relative to
the unpolarized continuum level $I_{\rm c}$ and per 1.8~\kms\ velocity bin) 
in the circular polarization profile produced by Least-Squares
Deconvolution (see Sec.~\ref{sec:obs}), while columns 6 indicates the 
rotational cycle associated with each exposure (using the ephemeris given 
by Eq~\ref{eq:eph}).  Exposure times are equal to $4\times600$~s for all 
observations. }   
\begin{tabular}{cccccc}
\hline
Date & HJD          & UT      & \sn\  & $\sigma_{\rm LSD}$ & Cycle \\
(2005) & (2,453,000+) & (h:m:s) &       &   (\ptt)  &  \\
\hline
Jun~19 & 540.86507 & 08:39:04 & 100 & 5.5 & 0.132 \\ 
\hline
Jun~20 & 541.80878 & 07:18:04 & 100 & 4.9 & 0.276 \\ 
Jun~20 & 541.83908 & 08:01:42 & 100 & 4.9 & 0.281 \\ 
Jun~20 & 541.86941 & 08:45:23 & 90 & 5.8 & 0.286 \\ 
\hline
Jun~21 & 542.81897 & 07:32:49 & 110 & 4.1 & 0.431 \\ 
Jun~21 & 542.84993 & 08:17:24 & 110 & 4.2 & 0.436 \\ 
Jun~21 & 542.88152 & 09:02:54 & 110 & 4.1 & 0.441 \\ 
\hline
Jun~22 & 543.89201 & 09:18:04 & 70 & 7.2 & 0.595 \\ 
Jun~22 & 543.92304 & 10:02:46 & 70 & 6.5 & 0.600 \\ 
Jun~22 & 543.95467 & 10:48:19 & 90 & 5.3 & 0.605 \\ 
\hline
Jun~23 & 544.87704 & 08:56:36 & 120 & 3.9 & 0.746 \\ 
Jun~23 & 544.90808 & 09:41:17 & 120 & 3.9 & 0.751 \\ 
Jun~23 & 544.93912 & 10:25:60 & 110 & 4.1 & 0.756 \\ 
\hline
Jun~24 & 545.88764 & 09:11:56 & 120 & 3.8 & 0.901 \\ 
Jun~24 & 545.91867 & 09:56:37 & 120 & 3.9 & 0.906 \\ 
Jun~24 & 545.94970 & 10:41:19 & 110 & 4.2 & 0.911 \\ 
\hline
Jun~25 & 546.87674 & 08:56:19 & 120 & 3.8 & 1.052 \\ 
Jun~25 & 546.90776 & 09:40:59 & 120 & 3.7 & 1.057 \\ 
Jun~25 & 546.93878 & 10:25:40 & 120 & 3.8 & 1.062 \\ 
\hline
Jun~26 & 547.88109 & 09:02:40 & 120 & 3.9 & 1.206 \\ 
Jun~26 & 547.91212 & 09:47:21 & 120 & 3.9 & 1.211 \\ 
Jun~26 & 547.94317 & 10:32:03 & 110 & 4.0 & 1.216 \\ 
\hline
\end{tabular}
\label{tab:log}
\end{table}

\section{Observations}
\label{sec:obs}

Spectropolarimetric observations of V2129~Oph were collected with ESPaDOnS  in 
2005 June. The ESPaDOnS spectra span the whole optical domain (from 370 to 
1,000~nm) at a resolving power of 65,000.  A total of 22 circular-polarisation 
sequences were collected on 8 consecutive nights, each sequence consisting of 
4 individual subexposures taken in different polarimeter configurations.   The
full journal of observations is presented in Table~\ref{tab:log}.   The
extraction procedure is described in \citet[][see also Donati et al., 2007,  in
prep., for further details]{Donati97} and is carried out with  {\sc
Libre~ESpRIT}, a fully  automatic reduction package/pipeline installed at CFHT
for optimal extraction of  ESPaDOnS unpolarised (Stokes $I$) and circularly
polarised (Stokes $V$) spectra.   The peak signal-to-noise ratios per 2.6~\kms\
velocity bin are about 100 on average\footnote{At the time of our
observations, ESPaDOnS suffered a 1.3-mag light  loss due to severe damage to
the external jacket of optical fibres linking the  polarimeter with the
spectrograph. This has since been repaired.}.     

Least-Squares Deconvolution (LSD; \citealt{Donati97}) was applied to all
observations.   The line list we employed for LSD is computed from an {\sc
Atlas9} LTE model atmosphere  \citep{Kurucz93} and corresponds to a K5IV
spectral type ($\teff=4,500$~K and  $\logg=3.5$) appropriate for V2129~Oph (see
Sec.~\ref{sec:par}). We selected only  moderate to strong spectral lines 
whose  synthetic profiles had line-to-continuum core  depressions larger than
40\% neglecting all non-thermal broadening mechanisms. We omitted the spectral
regions within strong lines not formed mostly in the photosphere, such as the
Balmer and He  lines, and the \caii\ H, K and infrared triplet (IRT) lines. 
Altogether, about 8,500 spectral features  are used in this process, most of
them from \fei. Expressed in units of the  unpolarised continuum level $I_{\rm
c}$, the average noise levels of the  resulting LSD signatures range from 3.7
to 7.2$\times10^{-4}$  per 1.8~\kms\ velocity bin. They are listed  in
Table~\ref{tab:log}.  

Zeeman signatures, featuring full amplitudes of about 0.5\%, are clearly 
detected in the LSD profiles of all spectra (e.g., see Fig.~\ref{fig:lsd}).  
Circular polarisation is also detected in most emission lines, and in
particular in  the 587.562~nm \hei\ line and in the \caii\ 
(IRT) known as good  tracers of magnetospheric accretion \citep{Johns99b,
Valenti04}.   To increase \sn, LSD profiles from data collected within each
night were averaged  together. Given the very small phase range (typically
1\%) over which nightly  observations were carried out, no blurring of the
Zeeman signatures is expected  to result from this procedure.  The resulting set
of LSD Stokes $I$ and $V$ profiles  (phased according to the ephemeris of
eqn.~\ref{eq:eph}) is shown in  Fig.~\ref{fig:pho} while that corresponding to
the \caii\ IRT\footnote{Note that  the 3 components of the \caii\ IRT were
averaged together in a single profile to  increase \sn\ further.} and \hei\
emission lines are shown in Figs.~\ref{fig:irt} and  \ref{fig:he}.   The
corresponding longitudinal fields, computed from the first order moment of the 
Stokes $V$ profiles \citep{Donati97}, are listed in Table~\ref{tab:beff}.  

\begin{figure}
\center{\includegraphics[scale=0.35,angle=-90]{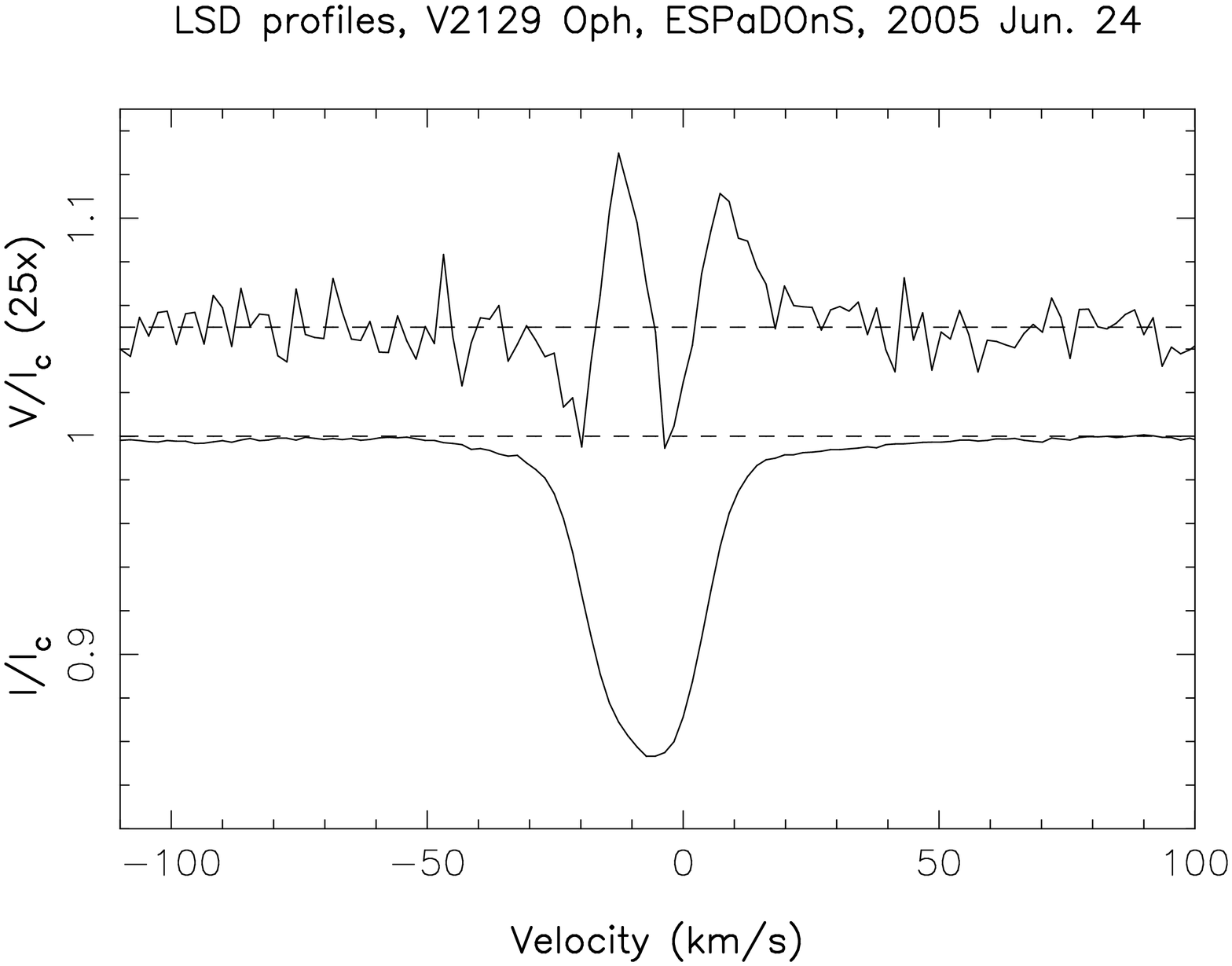}} 
\caption[]{LSD circularly-polarized and unpolarized
profiles of V2129~Oph (top, bottom curves respectively) on 2005 June~24 (phase 0.905).  
The mean polarization profile is expanded by a factor of 25 and shifted upwards 
by 1.05 for display purposes.  }
\label{fig:lsd}
\end{figure}

\begin{figure}
\center{\hbox{\includegraphics[scale=0.58,angle=-90]{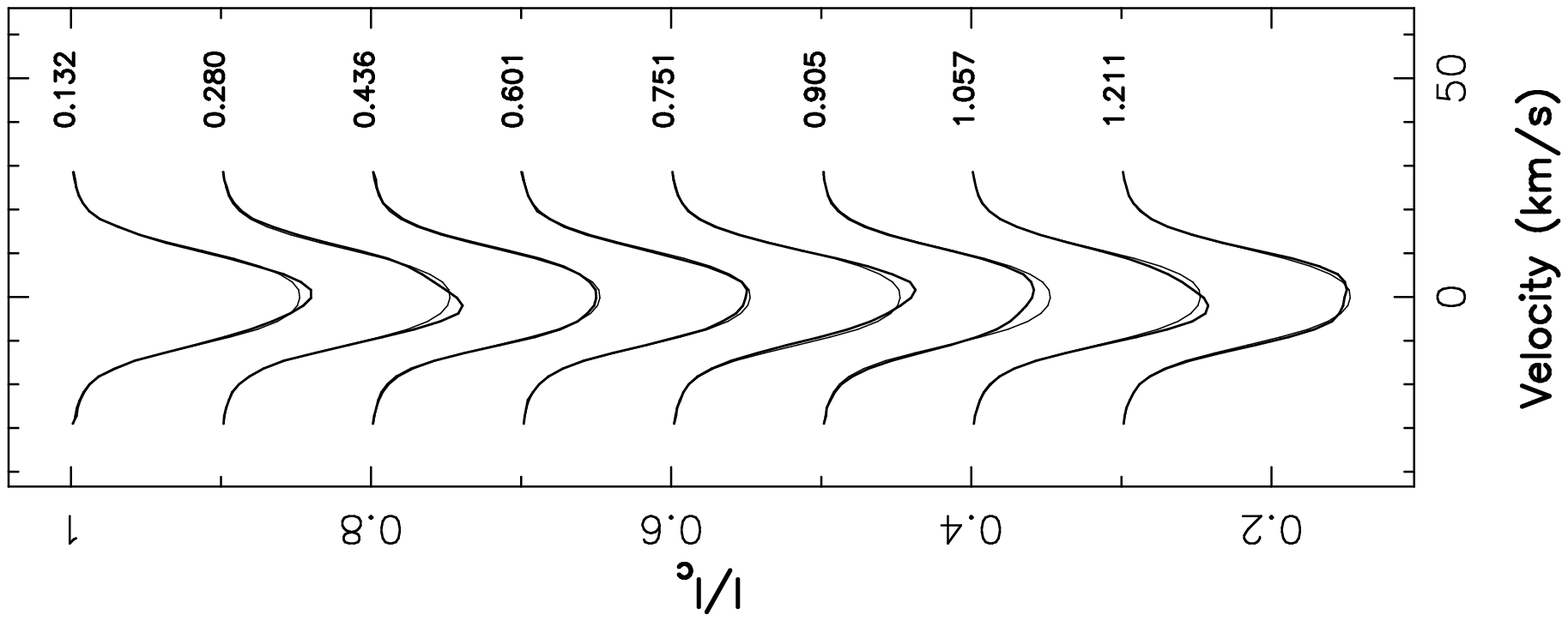}\hspace{1mm}
              \includegraphics[scale=0.58,angle=-90]{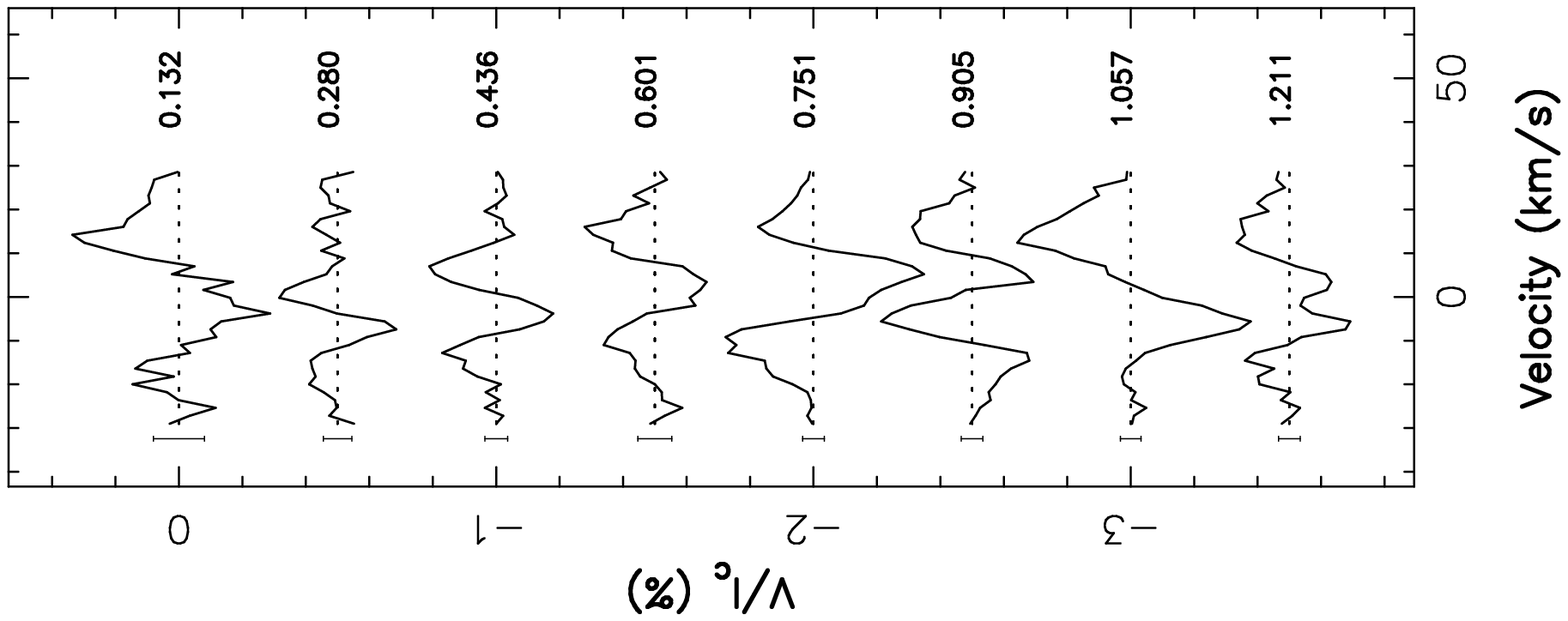}}} 
\caption[]{Average Stokes $I$ (left panel) and Stokes $V$ LSD profiles (right panel) 
of photospheric lines, for each observing night (top to bottom).  The mean Stokes $I$ 
LSD profile (averaged over the complete set, thin line) is also shown to emphasise 
temporal variations.  The rotational cycle associated with each observation is noted next
to each profile;  3$\sigma$ error bars are also shown to the left of each Stokes $V$ profile. 
All profiles are plotted in the velocity rest frame of V2129~Oph.  }
\label{fig:pho}
\end{figure}

\begin{figure}
\center{\hbox{\includegraphics[scale=0.58,angle=-90]{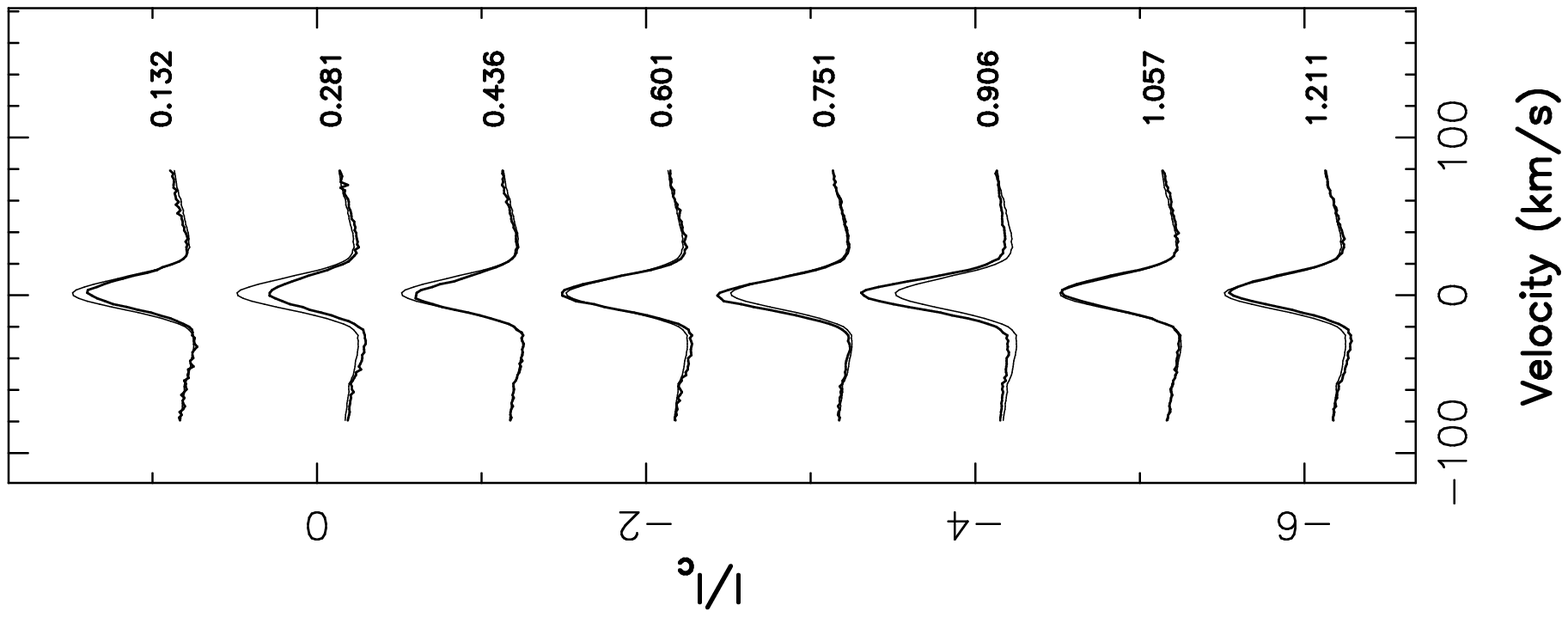}\hspace{1mm}
              \includegraphics[scale=0.58,angle=-90]{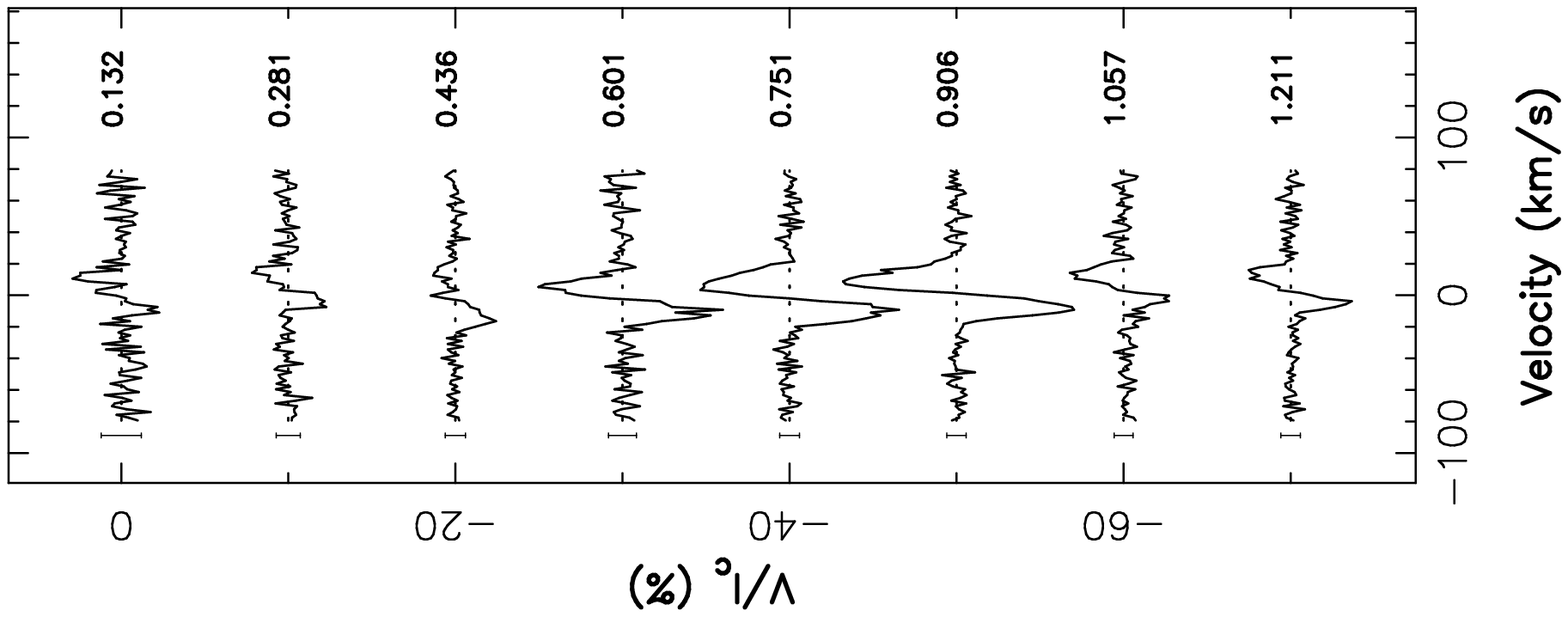}}}
\caption[]{Same as Fig.~\ref{fig:pho} for the \caii\ IRT lines.} 
\label{fig:irt}
\end{figure}

\begin{figure}
\center{\hbox{\includegraphics[scale=0.58,angle=-90]{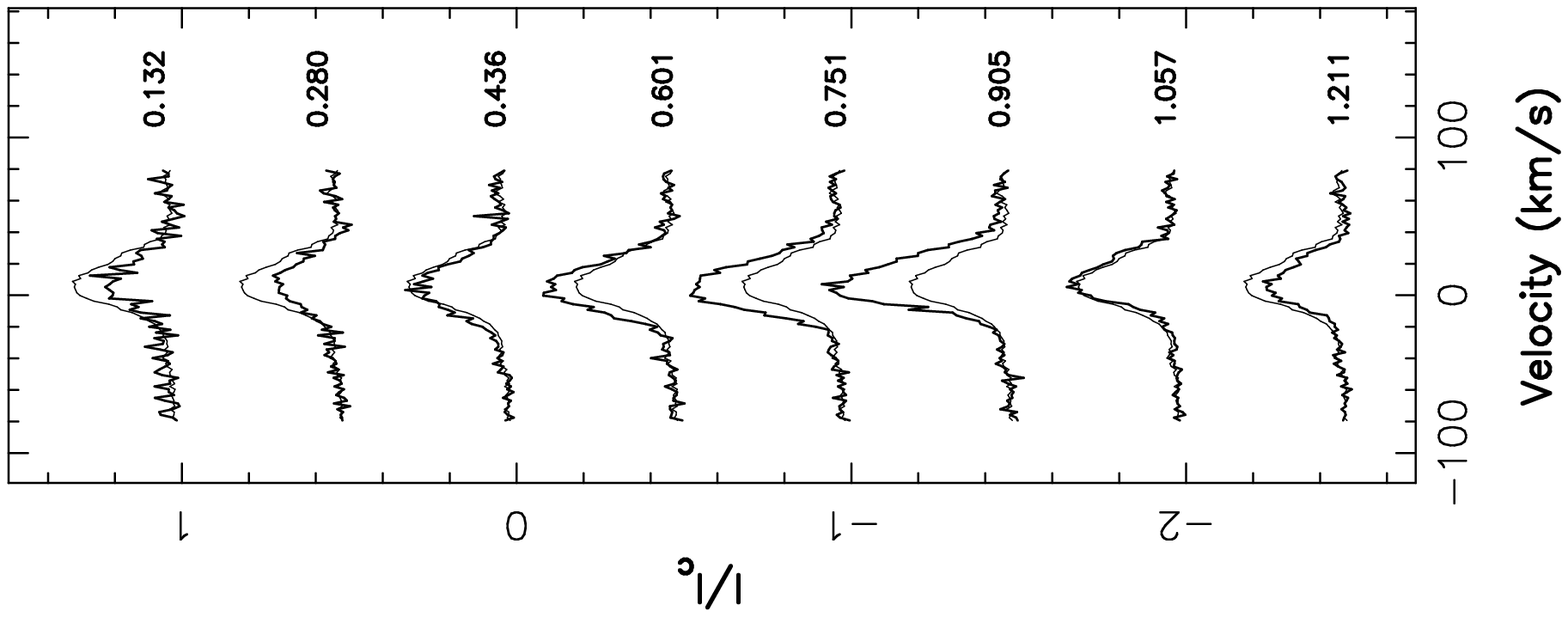}\hspace{1mm}
              \includegraphics[scale=0.58,angle=-90]{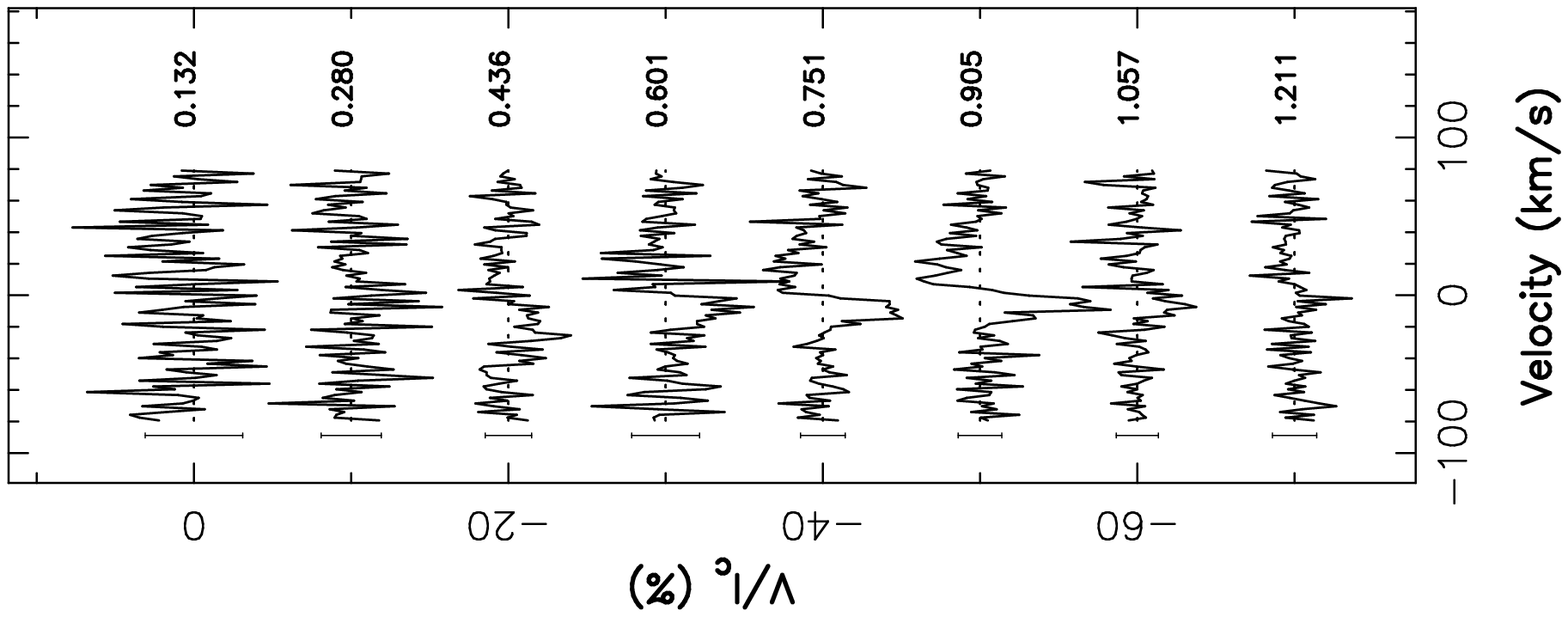}}}
\caption[]{Same as Fig.~\ref{fig:pho} for the \hei\ (right) lines.} 
\label{fig:he}
\end{figure}

\section{Rotational modulation}
\label{sec:zee}

In this section, we examine the temporal variations and shapes of  the Stokes
$I$ and $V$  LSD profiles and emission lines throughout the period of our
observations.  In  particular, we show that most of the temporal variations of
line profiles are  attributable to rotational modulation.  

\subsection{Photospheric lines and accretion proxies}

The  \caii\ IRT 
and \hei\ emission lines exhibit the strongest and simplest evolution over the 
period of our observations.  The fluxes and the amplitudes of the Zeeman
signatures in the emission lines are strongest on nights 5 and 6, and weakest
at the beginning  and end of our monitoring (see Figs.~\ref{fig:irt} and
\ref{fig:he}, right panels).   The typical timescale of this fluctuation is 
compatible with the rotation period of 6.53~d derived by
\citet{Shevchenko98}.   Longitudinal fields derived from both emission lines
(see  Table~\ref{tab:beff}) are plotted in Fig.~\ref{fig:blon} as a function
of  rotation cycle. These field estimates return to their initial values
(within  error bars) once the star has completed one full rotational cycle,
indicating that  the temporal fluctuations of the Zeeman signatures are mostly
attributable to  rotational modulation.  It also suggests that the  magnetic
fields are mostly monopolar  over the formation region of the Zeeman
signatures, which we identify with the accretion  spots at the footpoints of
accretion funnels. Very similar behaviour is  reported for other cTTSs
\citep[e.g.,][]{Valenti04}, further strengthening our  interpretation.  

\begin{table}
\caption[]{Longitudinal magnetic field of V2129~Oph, as estimated from LSD profiles 
($B_{\rm LSD}$, col.~2), the \caii\ IRT emission core ($B_{\rm IRT}$, col.~3) and 
the \hei\ line ($B_{\rm \hei}$, col.~4).  Column 7 lists the veiling parameter $r$ at each 
phase, defined as the relative decrease in the depth of photospheric lines (with respect 
to their average intensity) and estimated from LSD Stokes $I$ profiles.  
Rotational cycles (col.~1) are computed according to Eq~\ref{eq:eph}. }
\begin{tabular}{ccccc}
\hline
Cycle  & $B_{\rm LSD}$ & $B_{\rm IRT}$ & $B_{\rm \hei}$ & $r$ \\
       & (G) & (kG) & (kG) & \\
\hline 
0.132 & $-97\pm17$ & $0.23\pm0.07$ & $0.47\pm0.62$ & $+0.01$ \\
0.281 & $-6\pm9$   & $0.24\pm0.05$ & $0.71\pm0.39$ & $+0.02$ \\
0.436 & $-14\pm8$  & $0.33\pm0.03$ & $0.78\pm0.17$ & $ 0.00$ \\
0.601 & $-35\pm11$ & $0.44\pm0.04$ & $1.10\pm0.19$ & $-0.01$ \\
0.751 & $+51\pm7$  & $0.64\pm0.03$ & $1.11\pm0.09$ & $+0.02$ \\
0.906 & $-94\pm7$  & $0.67\pm0.02$ & $1.22\pm0.08$ & $-0.06$ \\
1.057 & $-180\pm7$ & $0.30\pm0.03$ & $0.91\pm0.15$ & $ 0.00$ \\
1.211 & $-39\pm7$  & $0.28\pm0.03$ & $0.72\pm0.22$ & $+0.02$ \\
\hline
\end{tabular}
\label{tab:beff}
\end{table}

The equivalent width of the \caii\ IRT line emission varies from 14 to 21~\kms\
(0.04 to 0.06~nm), while the \hei\ line emission varies from 6 to 18~\kms\
(0.012 to 0.036~nm).   We note that the variation of this emission strength is 
apparently not exactly periodic. While the \caii\ IRT and \hei\ emission 
strengths are equal to 15 and 6~\kms\ respectively at the beginning of our 
observations, they are slightly larger (by about 2 and 1~\kms\ respectively) 
one complete rotational cycle later.  This intrinsic variability of  the
emission strength is however significantly smaller  than the rotational
modulation itself.  

We suggest that the \caii\ IRT and \hei\ emission lines comprise two physically
distinct components.   We attribute the first of these, the accretion
component, to localised accretion spots at the surface of the star  whose
visibility varies as the star rotates, giving rise to rotational modulation of
the emission. A second, chromospheric component arises in basal  chromospheric
emission distributed more or less evenly  over the surface of the star, 
producing a time-independent emission component.  Assuming that the minimum
line  emission strength over the rotation cycle roughly corresponds to the
chromospheric  component, we obtain that the ratio of the accretion component
to the  basal chromospheric  emission is about 1:2 and 2:1 for the \caii\ IRT
and \hei\ lines respectively.   If we further assume that the chromospheric
component is mostly unpolarised while  Stokes $V$ signatures arise mainly in
the accretion component, we deduce that the Zeeman signatures are diluted
within the full line emission fluxes by factors of 1:3 and 2:3 for the \caii\ IRT
and \hei\ lines respectively.  The longitudinal field strengths derived from
each line need to be scaled up by  factors of 3:1 and 3:2 to recover the
longitudinal field corresponding to the  accretion component only.  In this way
we estimate the projected field strength within accretion spots at maximum 
visibility to be about 2~kG.  

\begin{figure}
\center{\includegraphics[scale=0.35,angle=-90]{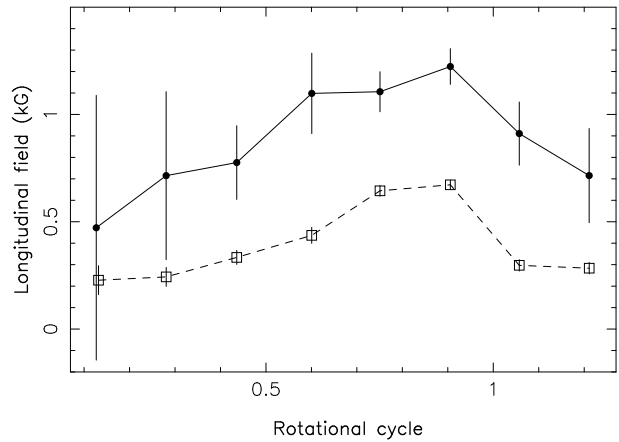}} 
\caption[]{Longitudinal field estimates derived from the Zeeman signatures of 
the \caii\ IRT (open squares) and \hei\ (filled dots) emission lines, as a function 
of rotational cycle; $\pm$1$\sigma$ error bars are also shown. }
\label{fig:blon}
\end{figure}

The Stokes $I$ and $V$ profiles of the LSD photospheric-line profiles also show
clear temporal variations (see  Fig.~\ref{fig:pho}). These are more complex
than those traced by the \caii\ IRT and \hei\  emission lines.  While the
Zeeman signatures of the emission lines are mostly  monopolar and keep the same
(positive) polarity throughout the whole rotation cycle, the photospheric
Stokes $V$  profiles exhibit  several sign switches across the rotation profile
(e.g., up to 3 at  phases 0.751 and 0.905)  and change polarity with time (see
Table~\ref{tab:beff}).   The longitudinal field strengths  associated with the
photospheric Zeeman signatures are usually less  than 100~G (except on rotation
cycle 1.057), much  weaker than those of the \caii\ IRT and \hei\ emission
lines.  This  indicates clearly that the underlying magnetic topology
associated with the photospheric Zeeman signatures is tangled and  mutipolar,
suggesting that emission and photospheric lines do not form over the  same
regions of the stellar surface.  Given the complexity of the photospheric 
Zeeman  signatures, a quick glance at the data cannot demonstrate conclusively
that  the observed modulation is consistent with pure rotational modulation.
We  can, however, verify that the shape of the Zeeman signature at the begining
of our  observations (phase 0.132 and 0.280) is similar to that observed one 
rotation later (phase 1.211).  Definite confirmation that this variability is 
entirely compatible with rotational modulation needs sophisticated stellar
surface  imaging tools;  this is achieved in Sec.~\ref{sec:zdi}.  

The photospheric Stokes $I$ LSD profile displays only moderate variability,
which is confined mainly to the line  core.  The  equivalent width of the
photospheric line profile varies by no more than a few percent (see
Table~\ref{tab:beff}), in agreement with previous published  reports on
V2129~Oph \citep{Padgett96, Eisner05}. This suggests that accretion spots on 
V2129~Oph are not hot enough to produce significant continuous ``veiling''
emission at the photospheric  level (and hence in LSD profiles), a reasonable
conclusion for a mildly accreting cTTS.   The photospheric Stokes $I$ profiles
exhibit shape variations reminiscent of the  distortions induced by cool
surface spots in spectral lines of active stars  \citep[e.g.,][]{Donati03}. 
Given the small amplitude of these distortions and the  relatively sparse
rotational sampling  of our data set, we again need stellar  surface imaging
tools to confirm that the profile distortions we detect are  compatible with
rotational modulation and attributable to surface spots; this  is presented in
Sec.~\ref{sec:dim}.  

\begin{figure*}
\center{\hbox{\includegraphics[scale=0.58,angle=-90]{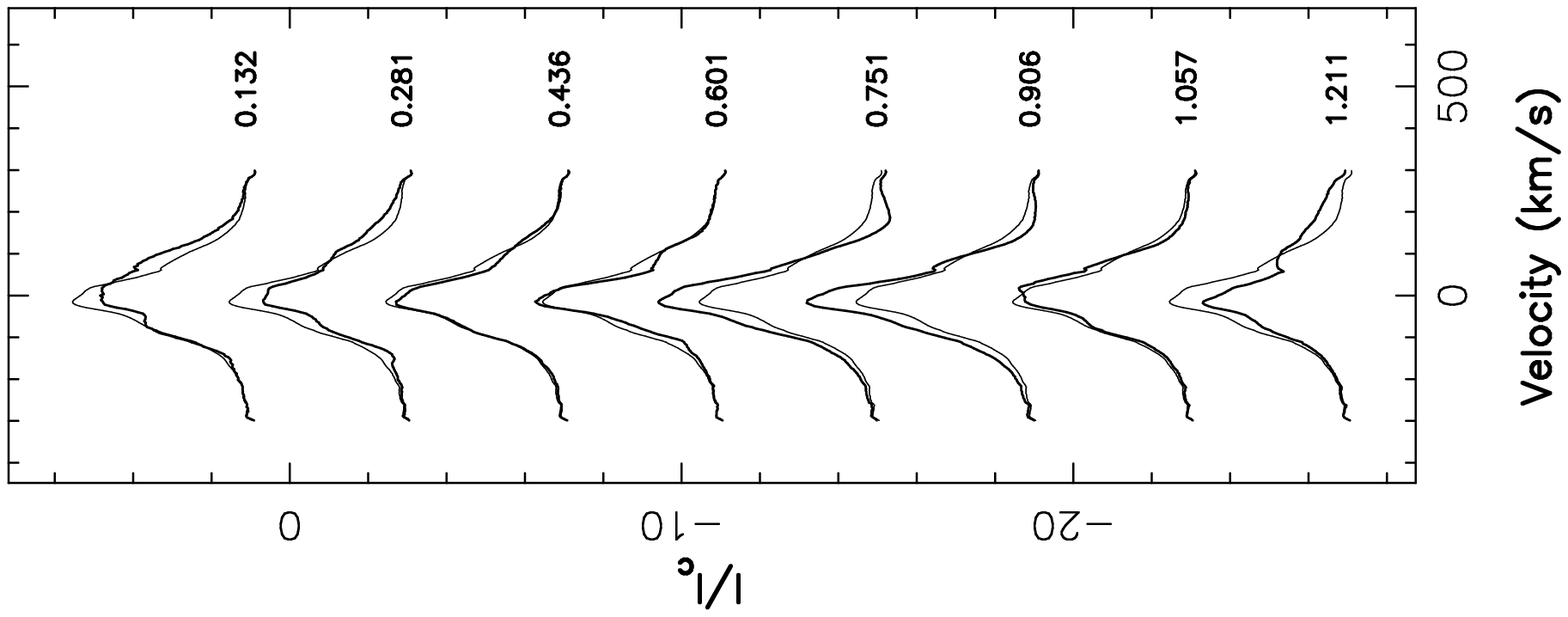}\hspace{1mm}
              \includegraphics[scale=0.58,angle=-90]{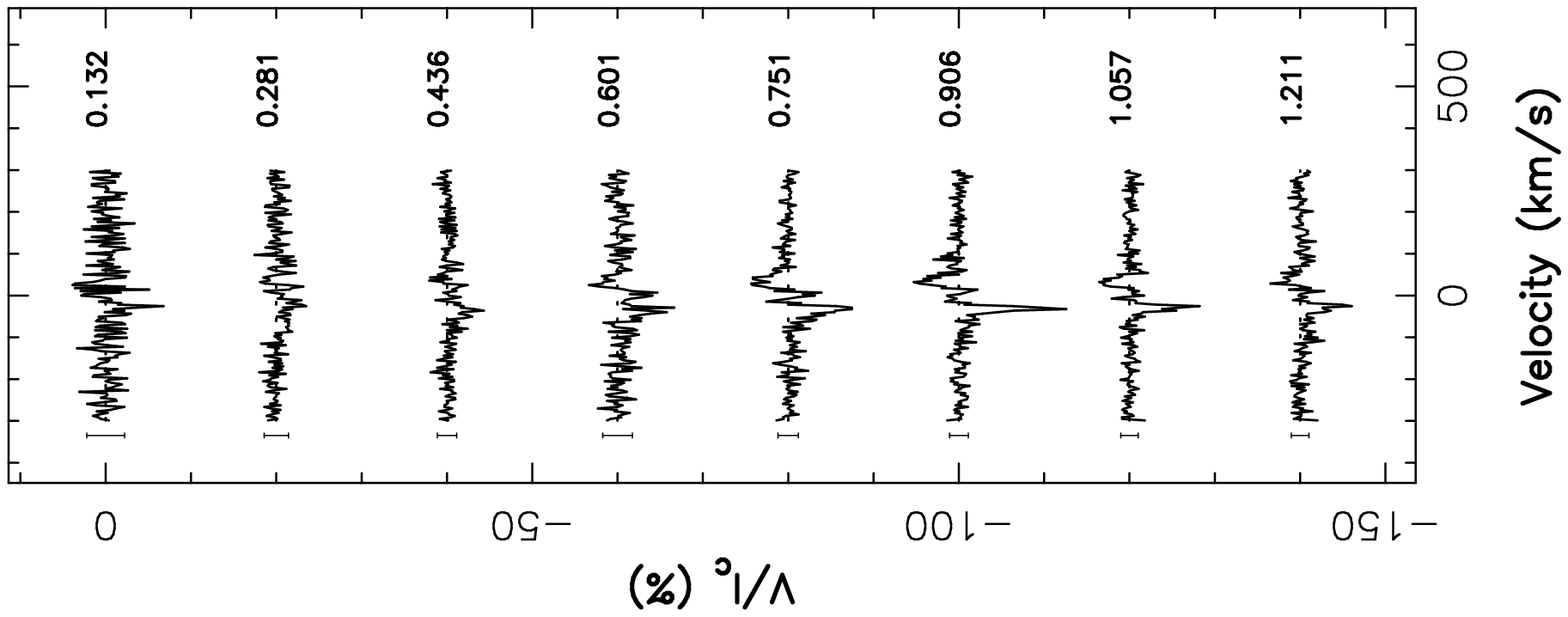}\hspace{4mm}
              \includegraphics[scale=0.58,angle=-90]{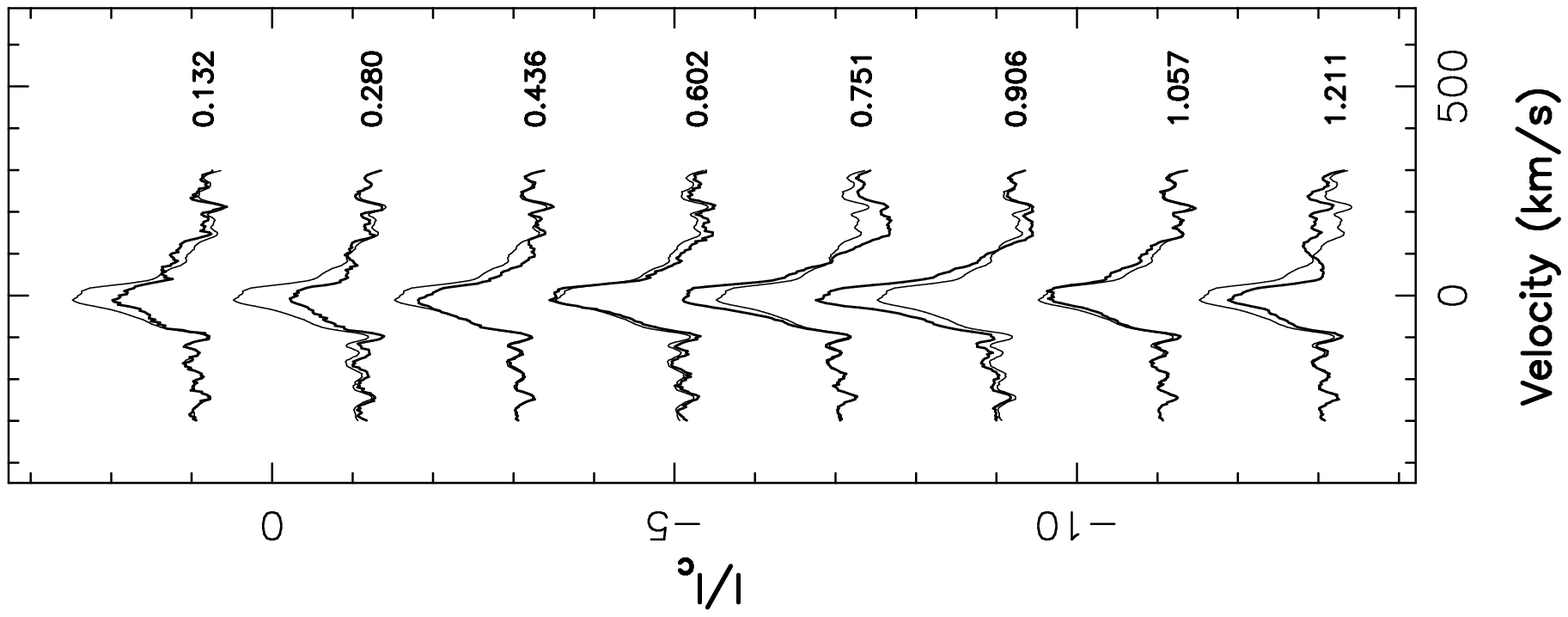}\hspace{1mm}
              \includegraphics[scale=0.58,angle=-90]{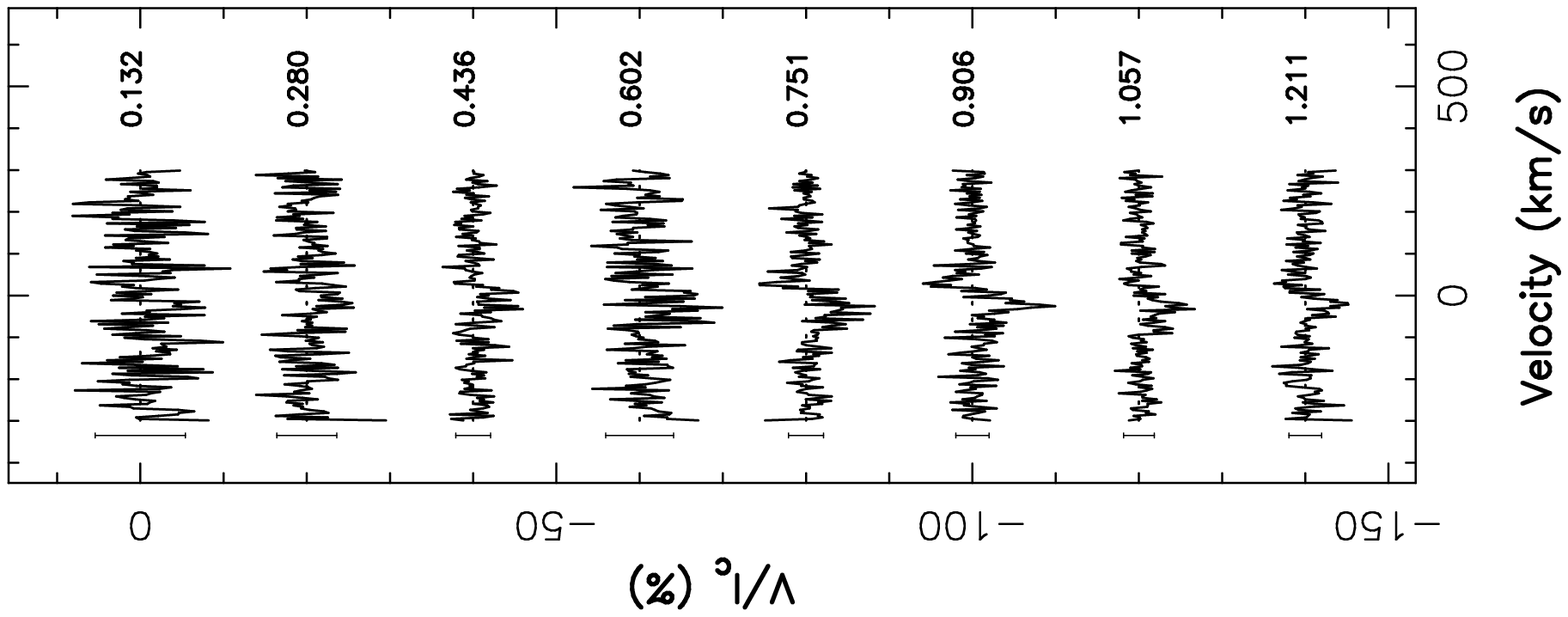}}}
\caption[]{Same as Figs.~\ref{fig:pho} for H$\alpha$ (left) and H$\beta$ (right). }
\label{fig:bal}
\end{figure*}

The average radial velocity of the photospheric lines is $-7.0$~\kms, which we
take  in the following to be the heliocentric radial velocity of the stellar
rest frame.   The \caii\ IRT emission core is centred at $-6$~\kms\ on average.
This is only very  slightly redshifted with respect to the stellar rest frame,
and  varies by less than $\pm1$~\kms\ over the rotational cycle (see
Fig.~\ref{fig:irt}).  The  width of the emission core (full width at half
maximum of 22~\kms) is comparable  to the rotational broadening of the
star.  

The \hei\ emission line centroid is more significantly redshifted relative to
the stellar rest frame than the \caii\ IRT emission  core, lying at 0~\kms\ on
average. Its full-width at half-maximum of 33~\kms\  is also significantly 
larger than that of the \caii\ emission core. Both discrepancies are partly
attributable  to the \hei\ line being composed of 6 different transitions,  the
last of which redshifted  from the main group by about 20~\kms.   The \hei\
centroid velocity varies in velocity by up to 2.5~\kms\  with rotation phase,
moving from blue to red between phase  0.45 and 1.05 and crossing the line
centre at phase 0.75 (see Fig.~\ref{fig:he}).   This is greater than the
velocity shifts of the \caii\ IRT emission core. It is qualitatively consistent
with the simple two-component model of emission lines  proposed above and
suggests that rotational modulation is mostly caused by the  accretion
component, which is  stronger in the \hei\ line than in the \caii\ IRT emission
line.  

The widths and redshifts of the \caii\ and \hei\ Zeeman signatures are similar
to  those of the unpolarised emission profiles. The Stokes $V$ profiles of the
\caii\ IRT  emission cores are mainly antisymmetric with respect to the line
centre. The  Zeeman  signature of the \hei\ line departs significantly  from
antisymmetry, with  a blue lobe both stronger and narrower than the red lobe.
This suggests that the \hei\  line forms in a region featuring non-zero
velocity gradients.

\begin{figure*}
\center{\hbox{\includegraphics[scale=0.25,angle=-90]{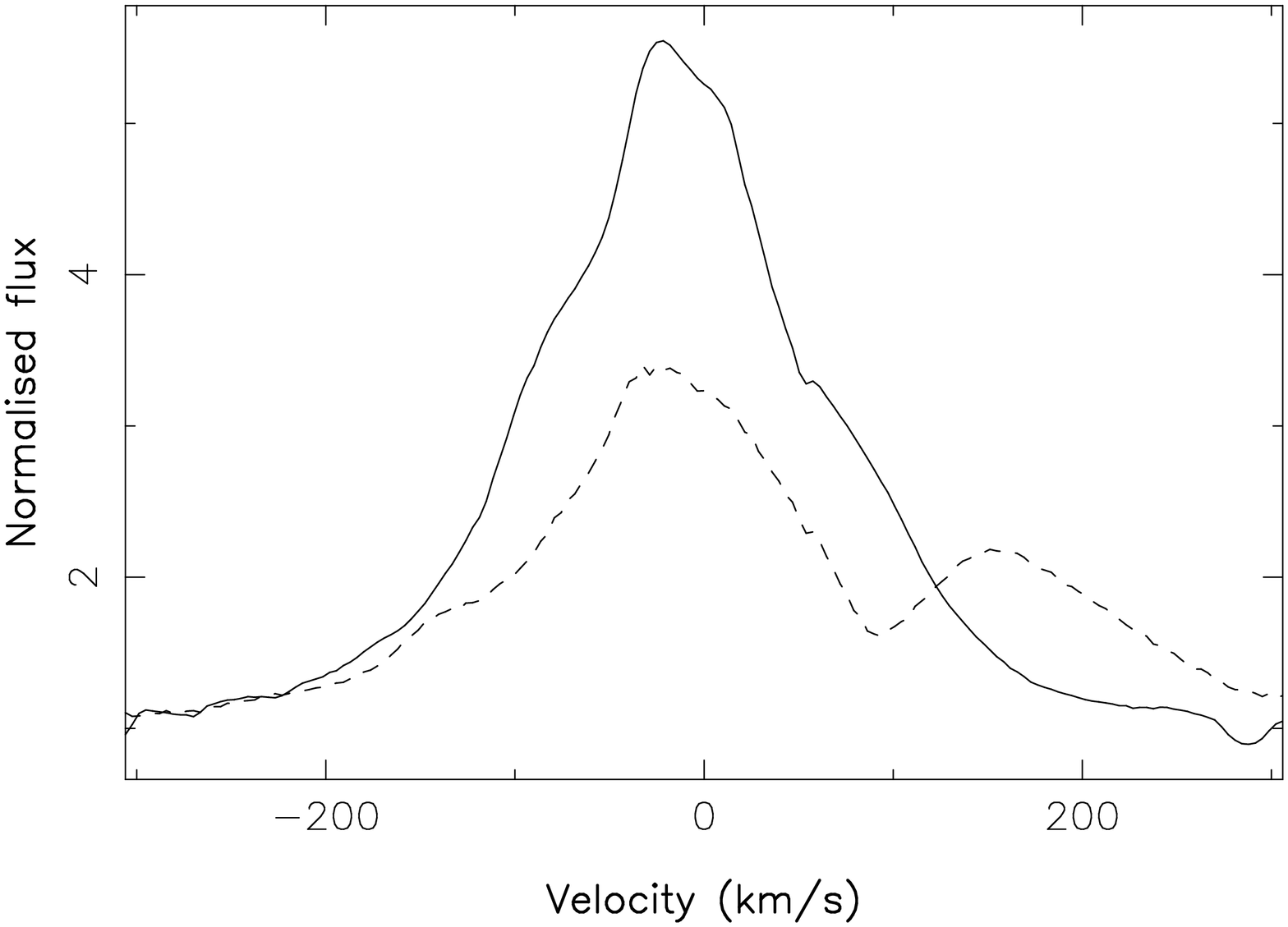}\hspace{1mm}
              \includegraphics[scale=0.25,angle=-90]{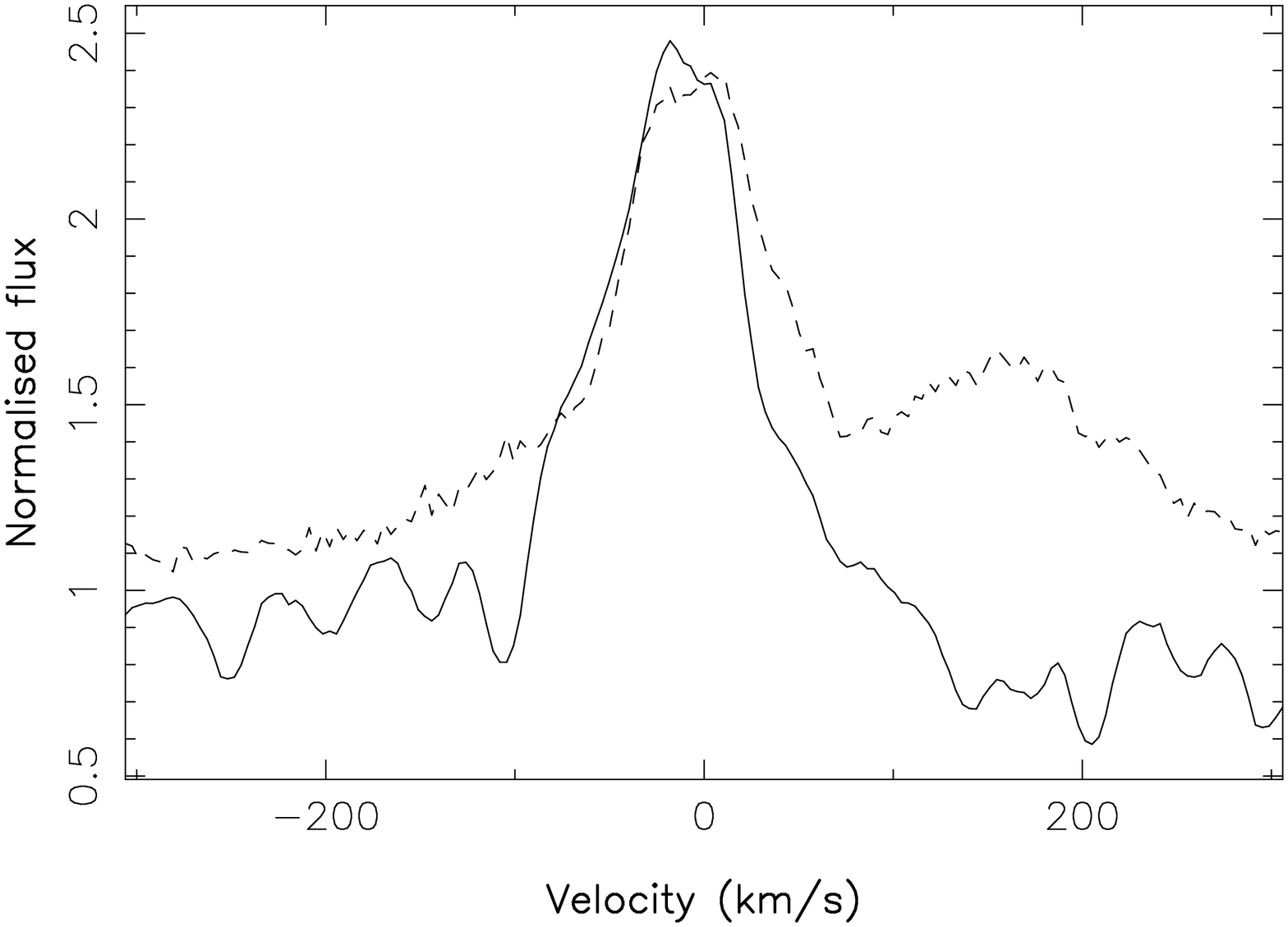}\hspace{1mm}
              \includegraphics[scale=0.25,angle=-90]{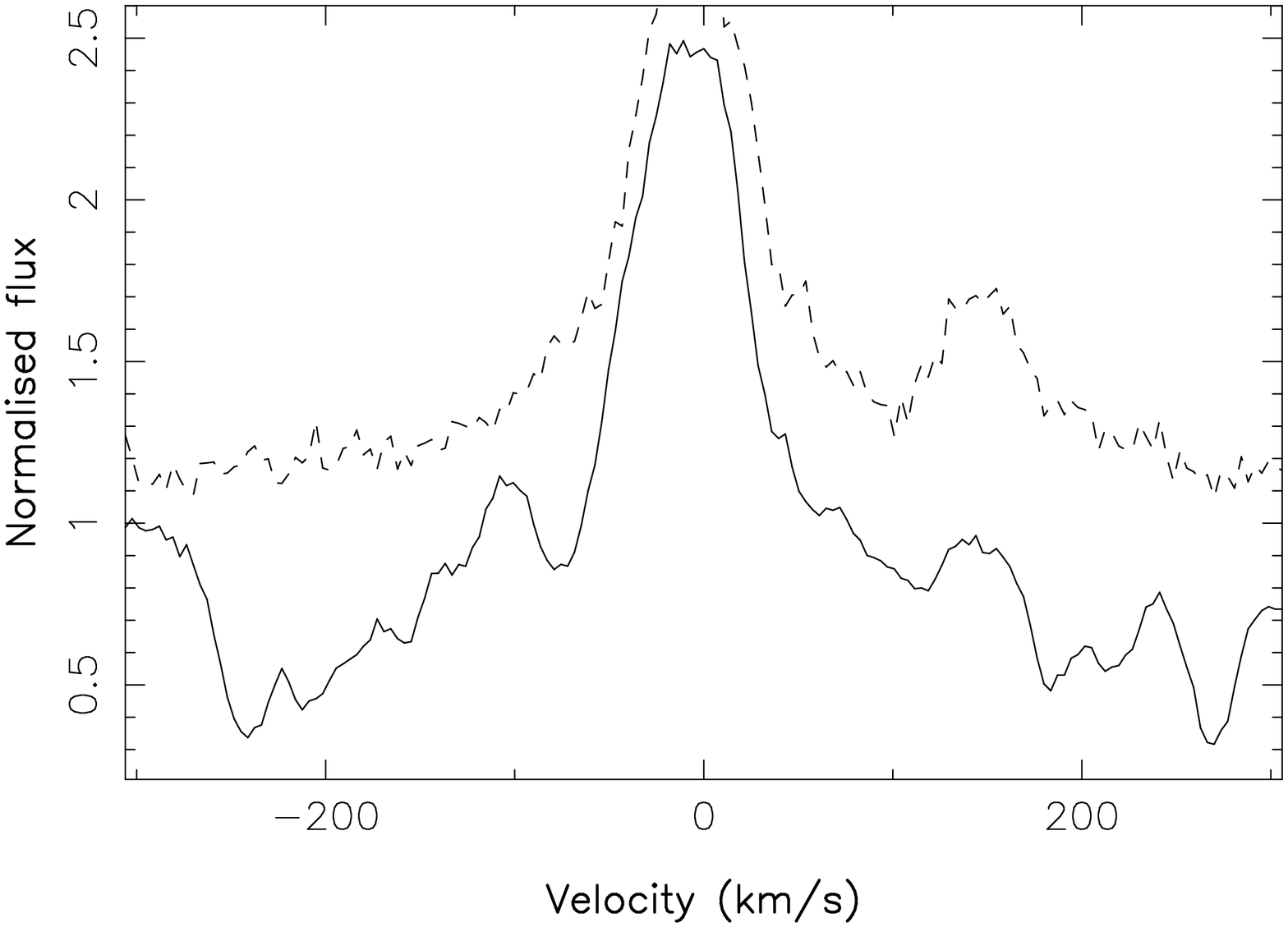}}}
\vspace{2mm}
\center{\hbox{\includegraphics[scale=0.32,angle=-90]{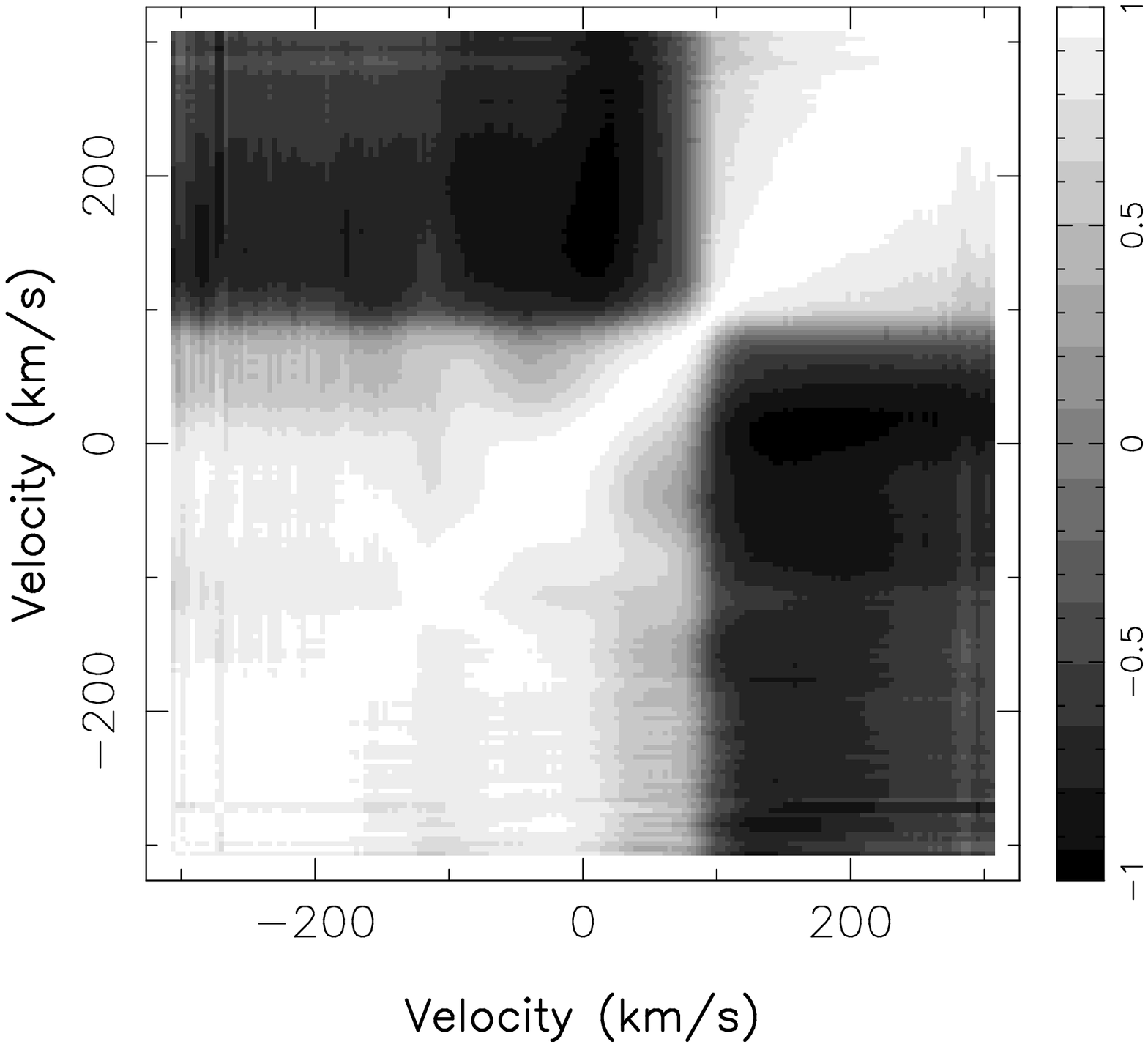}\hspace{1mm}
              \includegraphics[scale=0.32,angle=-90]{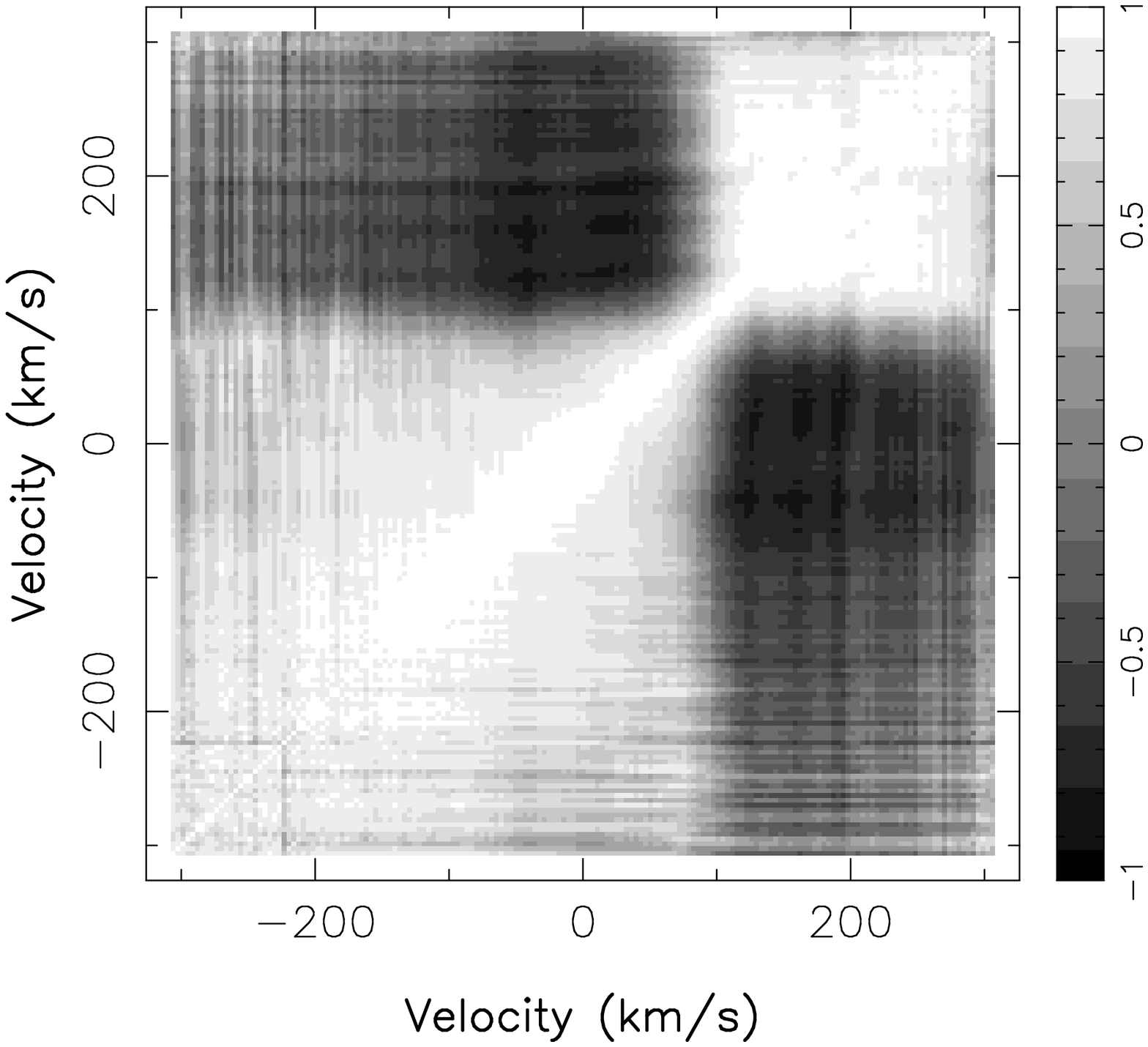}\hspace{1mm}
              \includegraphics[scale=0.32,angle=-90]{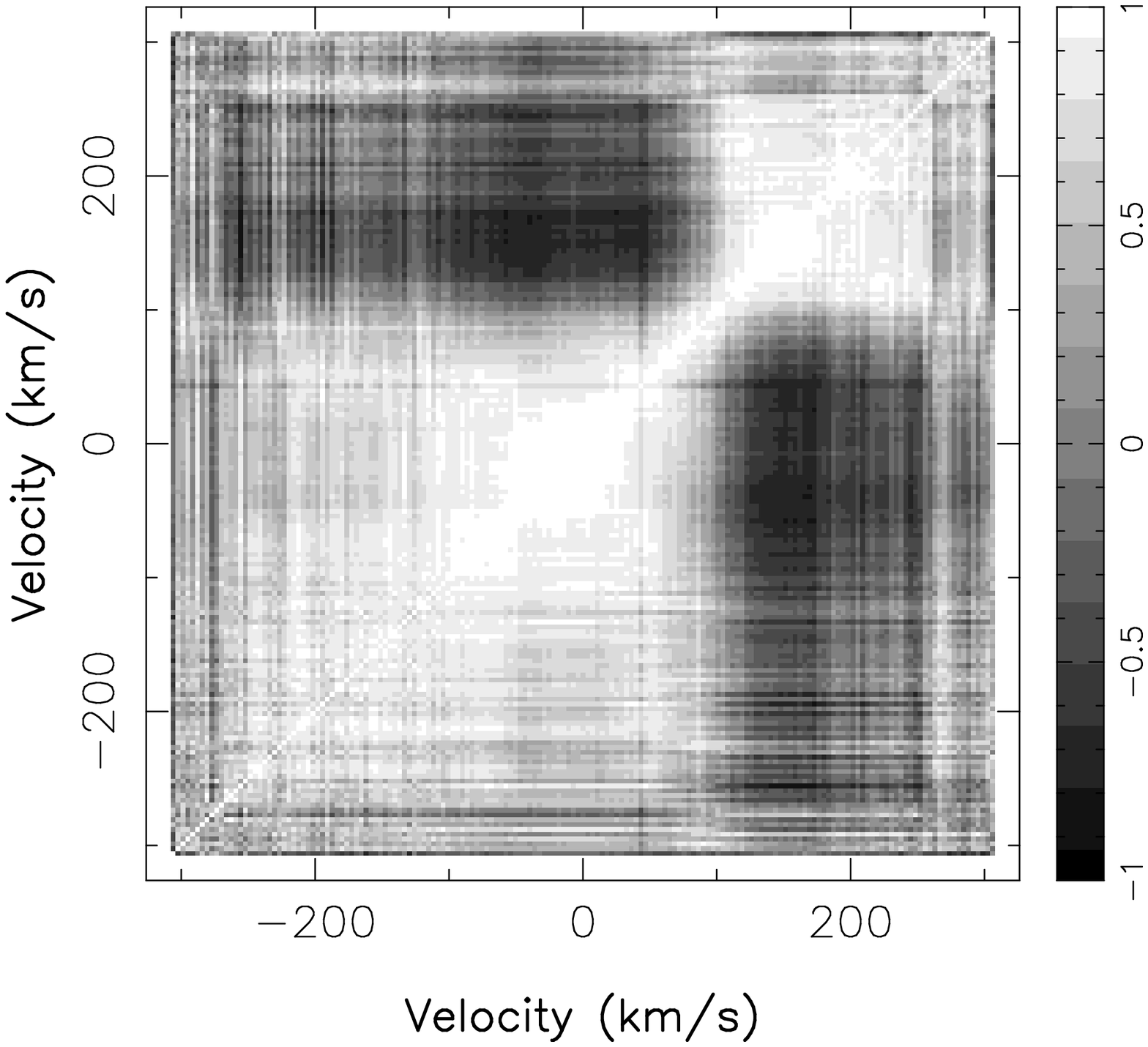}}}
\caption[]{Mean H$\alpha$ (left), H$\beta$ (middle) and H$\gamma$ (right) profiles (top panel,
full line) and associated standard deviations (top panel, dashed line), with corresponding
autocorrelation matrices (bottom panel).  Note that the standard deviation profiles are
multiplied by 3 and shifted upwards by 1 for display purposes.  In the lower panels,
white indicates perfect correlation and black means perfect anticorrelation.  }
\label{fig:bal2}
\end{figure*}

\subsection{Balmer lines}

H$\alpha$ and H$\beta$  lines exhibit strong emission with average
equivalent widths of 900 and 200~\kms, (i.e.,  2 and 0.32~nm) respectively.
They exhibit clear rotational modulation and  convey information that \caii\
IRT and \hei\ emission lines do not contain (see  Fig.~\ref{fig:bal}).  One
obvious difference is that their unpolarised emission  profiles are
significantly broader (with full widths at half maximum of 190 and  90~\kms\
for H$\alpha$ and H$\beta$ respectively) than their Zeeman signatures 
(70~\kms\ wide).  They also include a high-velocity component in their red wings 
(between $+100$ and $+300$~\kms) that alternately appears in absorption and 
emission (at cycles 0.75 and 1.21 respectively) but does not show up in the
Stokes $V$ profile.  Both points are readily visible on the average emission
and standard  deviation profiles of Balmer lines (see top panel of
Fig.~\ref{fig:bal2} for  H$\alpha$ to H$\gamma$).  

Autocorrelation matrices of Balmer emission profiles (see bottom panel of 
Fig.~\ref{fig:bal2}) reveal that the high-velocity redshifted component is 
anticorrelated with the central emission peak.  We propose that the central 
Balmer emission peak, which varies in phase with the \caii\ and \hei\ emission,
is partly  produced in the accretion shock. From its significant blueshift
(7~\kms\ with  respect to the stellar velocity rest frame for H$\alpha$ and
H$\beta$), the Balmer emission  likely includes a wind component as well,
strongest in H$\alpha$.  The anti-correlated modulation of the high-velocity  
redshifted component points to a different origin;  we suggest it is due to  
free-falling material in the preshock region of the accretion funnels.

\section{Active regions and accretion spots}
\label{sec:dim}

From the overview of the collected emission-line data in Sec.~\ref{sec:zee}, we
infer that rotational modulation is responsible for  most of the observed
variability in the  \caii\ IRT and \hei\ emission lines.   We demonstrate in
the present and following  sections that this is also the case for the  Stokes
$I$ and $V$ LSD profiles of  photospheric lines.   As a first step, we use
Doppler tomography to reconstruct the locations of dark spots on the surface of
V2129~Oph from  the time series of unpolarised photospheric line profiles. We
also  apply this Doppler imaging technique to the emission-line profiles, using
them as accretion proxies to determine where the accretion  spots are
located.  

\begin{figure}
\center{\hbox{\includegraphics[scale=0.58,angle=-90]{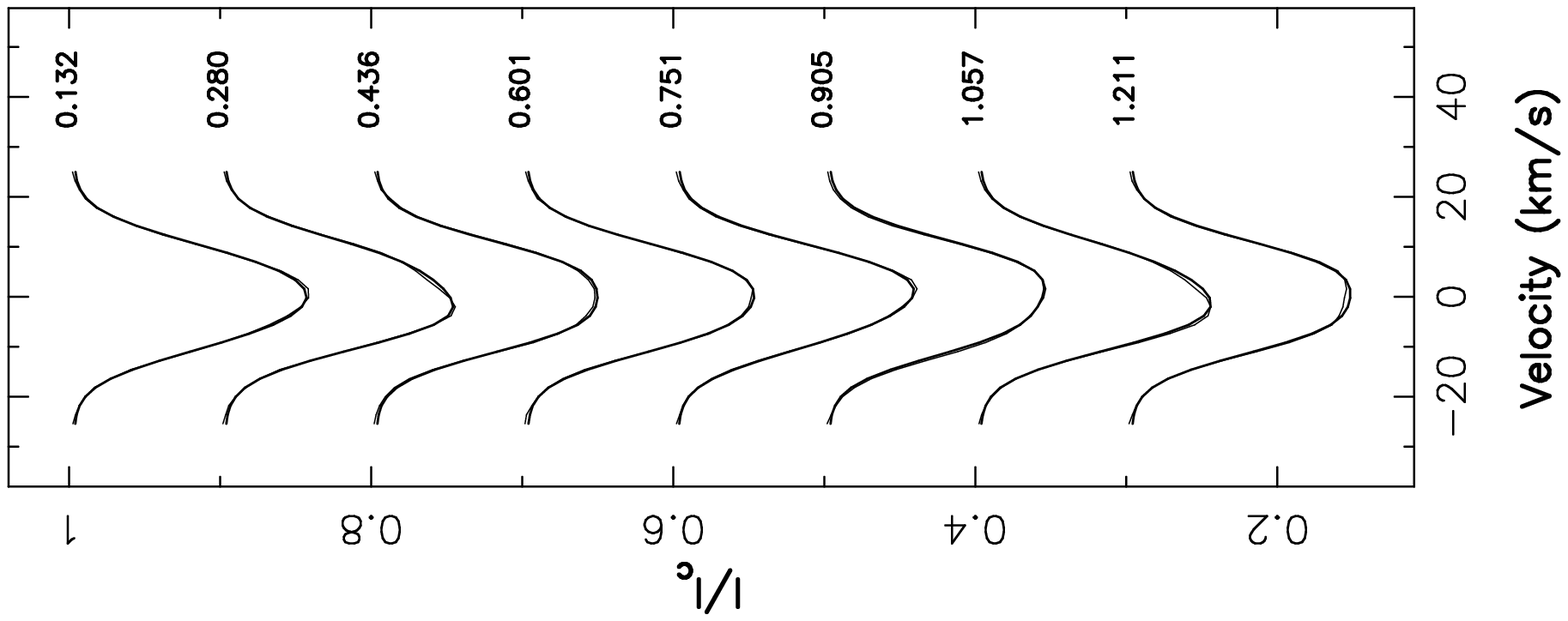} 
             \includegraphics[scale=0.58,angle=-90]{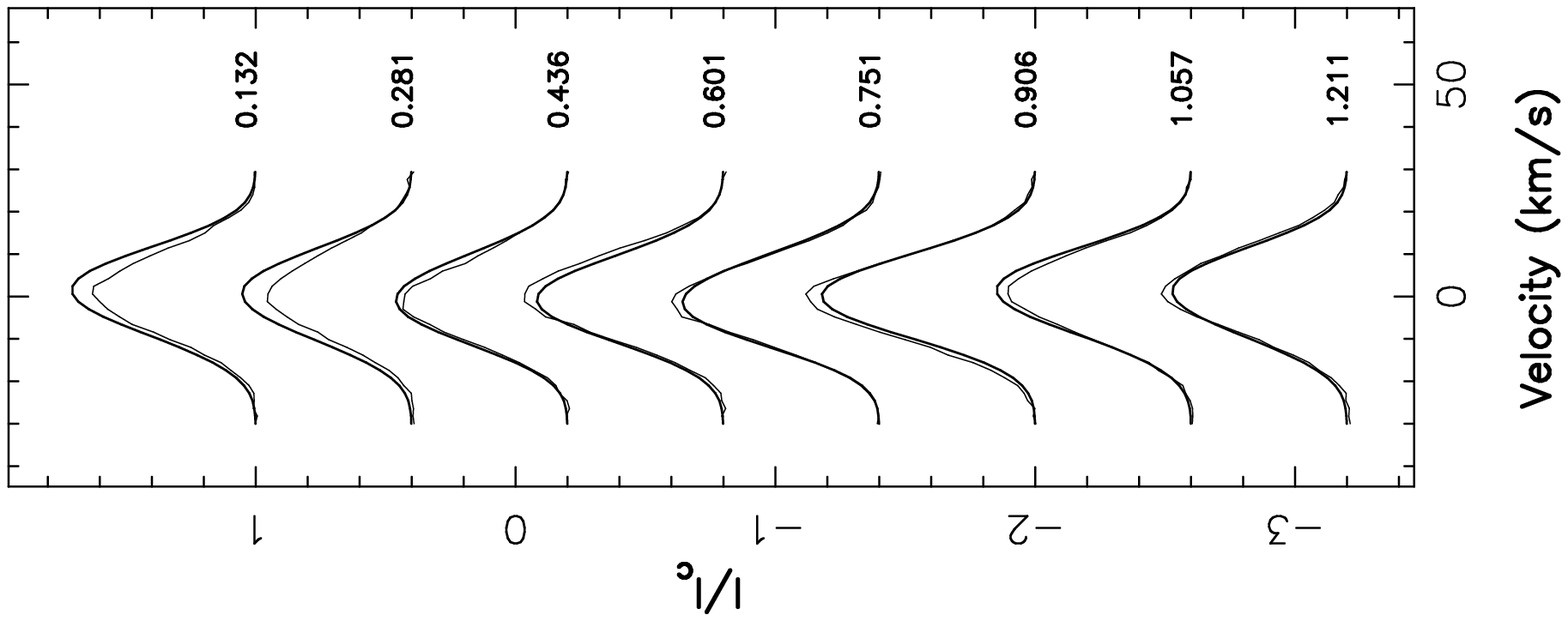}}} 
\caption[]{
Observed LSD Stokes $I$ (left panel) and \caii\ IRT emission core (right 
panel) of V2129~Oph (thin line) along with the maximum entropy fit to the data
(thick  line).  The rotational phase and cycle of each observation is written
next to each  profile. 
}  
\label{fig:fiti}
\end{figure}

\begin{figure*} 
\center{\hbox{\includegraphics[scale=0.50]{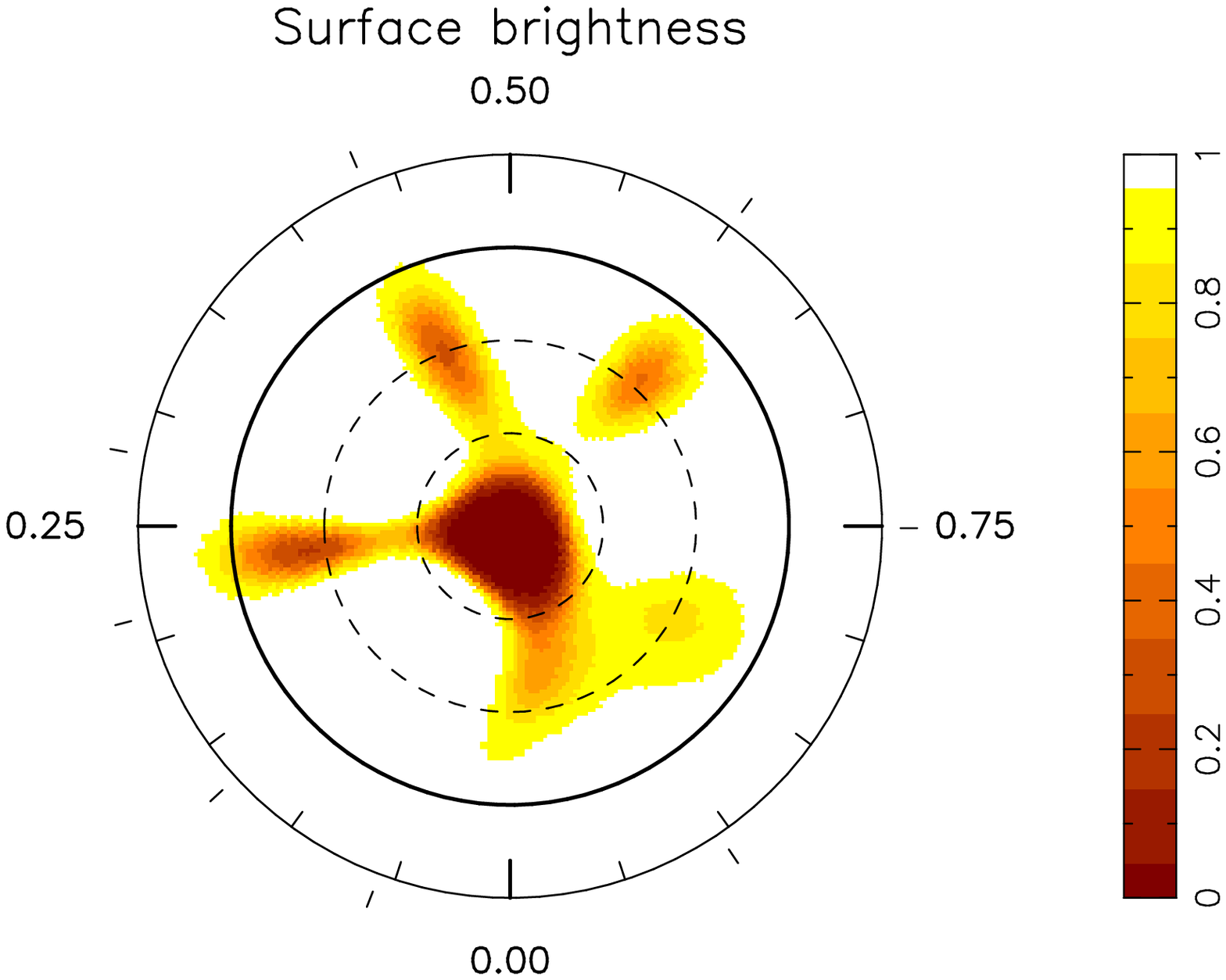}
              \includegraphics[scale=0.50]{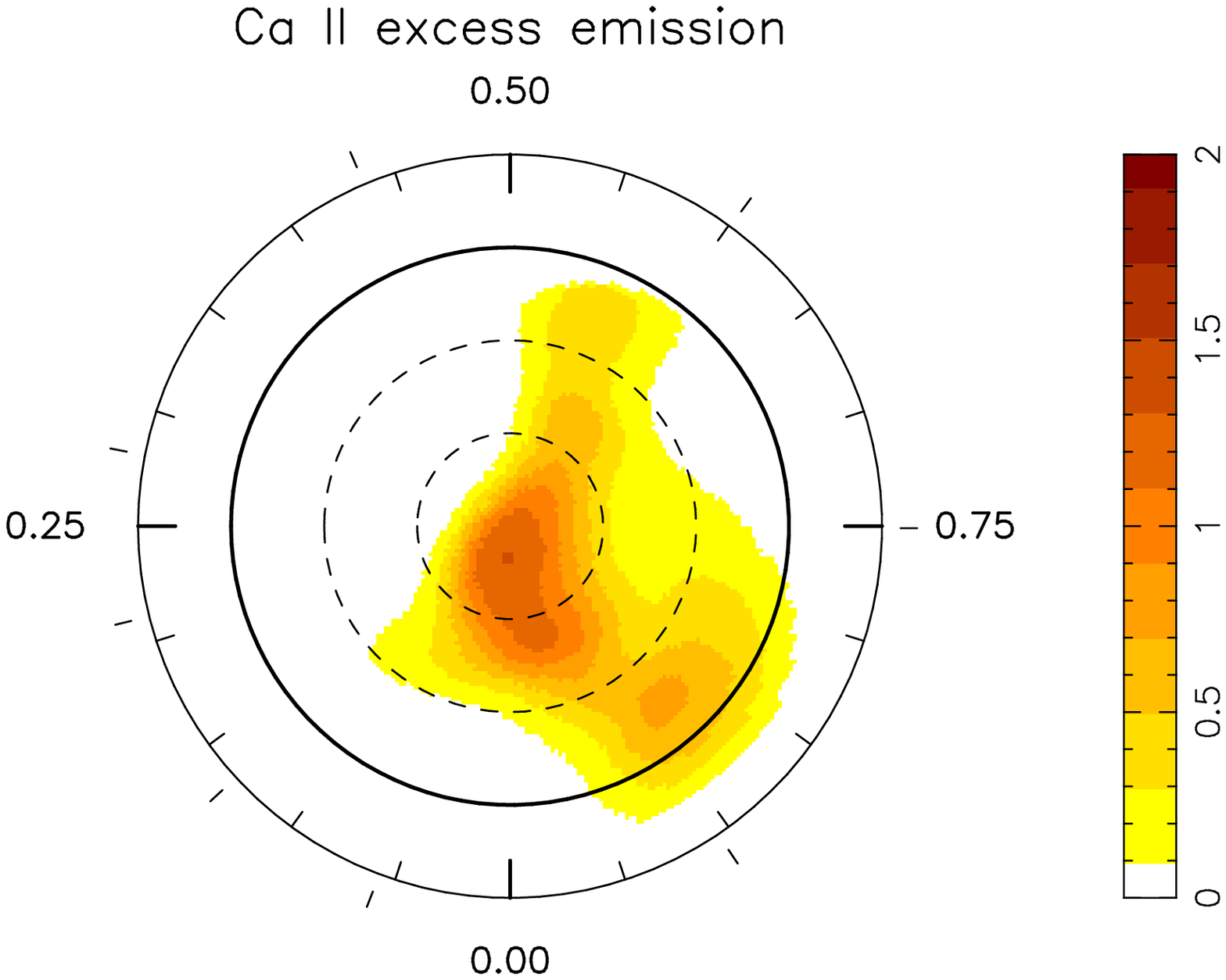}}} 
\caption[]{
Surface maps of the local surface brightness relative to that of the quiet
photosphere (left panel) and the local \caii\ excess emission relative to the
basal chromospheric emission (right panel)  on V2129~Oph, reconstructed from
our series of ESPaDOnS LSD Stokes $I$ and  \caii\ IRT emission line profiles
(Fig.~\ref{fig:fiti}).   The star is shown in flattened polar projection down
to latitudes of  $-30\degr$, with the equator depicted as a bold circle and
parallels as dashed circles.   Radial ticks around each plot indicate the
phases of observation.  
}
\label{fig:map}
\end{figure*}

\subsection{Dark spots}

As we noted in Sec.~\ref{sec:zee}, the distortions seen in the Stokes $I$  LSD
profiles of V2129~Oph resemble those caused by cool photospheric features  on
the surfaces of active stars. \citet{Shevchenko98} reach a similar 
conclusion from multicolour photometry.  In the following, we 
model the observed LSD profile  variations using tomographic
imaging, assuming
that dark surface features  are present at the surface of V2129~Oph.
We thus demonstrate  that the temporal variations in the LSD profiles
are mainly attributable to rotational  modulation, and 
determine where dark spots are located on the  surface of V2129~Oph.  

We remove the veiling  by simply scaling all profiles to the same  equivalent
width,  and use the maximum entropy image reconstruction code  of
\citet{Brown91} and \citet{Donati97b} to adjust the surface image iteratively
in order to optimise the fit to the series of Stokes $I$  profiles.  The image
quantity we reconstruct is  the local surface brightness (relative to the quiet
photosphere), varying from 1 (no spot)  to 0 (no light).  The local profile,
assumed to be constant over the whole star,  is obtained by matching a
Unno-Rachkovski profile \citep{Landi04} with no  magnetic fields to the LSD
profile of a K5 template star (61~Cyg) computed with  the same line list as
that used for V2129~Oph.   The synthetic profiles we finally obtain  (see 
Fig.~\ref{fig:fiti}, left panel) fit the observed Stokes $I$ LSD profiles at 
a level of about 0.15\% or \sn=700 and confirms that the profile distortions 
are attributable to cool spots at the surface of the star.  

Given the relative sparseness of our data set, one may argue that the success 
at fitting the Stokes $I$ LSD profiles is no more than a coincidence and does
not  truly demonstrate that the profiles are rotationally modulated.  To check
this,  we varied the assumed rotation period of the star over a small interval
about the  value of \citet{Shevchenko98} and reconstructed brightness images
for each of the  assumed rotation periods. We anticipate that the reconstructed
image will show minimal spot coverage at the correct rotation period  when the
data are fitted at a fixed \chisq, (or equivalently will yield the best fit  to
the data for a given spot coverage), if the profile variability is truly due to
rotational modulation.  This  technique has also been used to derive
constraints on the amount of differential  rotation shearing the photosphere of
V2129~Oph \citep{Donati00, Petit02, Donati03b}.   Assuming solid body rotation, the
optimal period we find is $6.56\pm0.02$~d,  compatible with the value of 6.53~d
derived by \citet{Shevchenko98}.   Assume now that the surface of the star
rotates differentially, with the  rotation rate varying with latitude $\theta$
as $\omeq-\dom\sin^2 \theta$,  \omeq\ being the angular rotation rate at the
equator and \dom\ the difference  in angular rotation rate between the equator
and pole.  We then find that optimal  fits to the Stokes $I$ data are achieved
for $\omeq=0.964\pm0.003$~\rpd\ and  $\dom=0.054\pm0.019$~\rpd, corresponding to
rotation periods of 6.52 and 6.90~d  for the equator and pole respectively. 
The error bar on \dom\ is too large  to detect unambiguously differential
rotation at the surface of the star;  we can  however exclude values of \dom\
that are either smaller than 0 or larger  than 0.1~\rpd\ (i.e., about twice the
shear at the surface of the Sun).   As a by-product, we very clearly
demonstrate that the temporal variability  of Stokes $I$ LSD profiles is due to
rotational modulation.  

The resulting map (see Fig.~\ref{fig:map}, left panel) includes one cool spot
close to the  visible pole and several additional ones at low to intermediate
latitudes.  The main  polar spot is undoubtedly real.  The low-latitude
appendages at phases 0.23 and 0.95  are also well constrained by observed
profiles featuring obvious line asymmetries.   The two features centred on
phase 0.44 and 0.62 are only associated with low level  line asymmetries and
their reality is less well established.  The total area covered  by
photospheric cool spots in the reconstructed image is about 7\%, with about
5\%  for the main polar spot only.  We emphasize that this total spot coverage
should be regarded as a lower limit.  Doppler imaging recovers successfully the
most significant brightness features,  such as the cool polar spot  seen here.
Even if the stellar surface is peppered with spots much smaller than our
resolution limit \citep[e.g.,][]{Jeffers06},  such features will be suppressed,
especially  in stars with moderate  rotation velocities.  

We find that $\vsini=14.5\pm0.3$~\kms\ and $\vrad=-7.0\pm0.2$~\kms\  at the
time of our observations, in perfect agreement with the recent published  
estimates ($15.2\pm0.9$ and $-7.17\pm0.25$~\kms, \citealt{Eisner05}).  Given
the  spectral resolution of ESPaDOnS (of about 5~\kms), the longitudinal
resolution of  the imaging process at the stellar equator is about 20\degr\ or
0.05 rotation cycle. This is consistent with the longitudinal extent of the
brightness features we  recovered.  We find that the information  content of
the brightness image is minimised for an axial inclination angle $i=45$\degr,
in agreement  with expectations from  spectrophotometric estimates (see
Sec.~\ref{sec:par}).  

\subsection{Accretion spots}

We proceed in a similar way to model the fluctuations of the \caii\ IRT
emission  profiles and to retrieve information about the location of accretion
spots at the  surface of V2129~Oph.  Thanks to their much smaller noise level,
narrower width and  simpler Zeeman signatures, and despite their weaker amount
of rotational modulation, the \caii\ emission lines are a better choice than
the \hei\ line for this modelling task.   The modelling accuracy is, however,
limited by the intrinsic variability of the unpolarised emission lines.  For
this modelling, we use the simple two-component  model described in
Sec.~\ref{sec:zee}, involving a basal chromospheric emission  component evenly
distributed over the surface of V2129~Oph and an accretion emission  component
arising in local accretion spots.  

We now reconstruct an image of the local \caii\ excess emission from  accretion
spots relative to the basal chromospheric emission.  We assume that the  \caii\
minimum emission profile (at phase 0.28, see Fig.~\ref{fig:irt}) provides a 
rough estimate of the basal chromospheric component.  To model this
chromospheric  component, we use a Gaussian local profile with a full width at
half maximum of  12~\kms\ and an equivalent width of 15~\kms;  given the
\vsini\ estimate derived  from Stokes $I$ LSD profiles, this model provides a
reasonable match to the minimum  emission profile at phase 0.28.  We use the
same local profile to model the  accretion component.  The fit we obtain is
shown in Fig.~\ref{fig:fiti} (right  panel) while the reconstructed map is
presented in Fig.~\ref{fig:map} (right  panel).  As a result of intrinsic
variability, the quality of the fit is poorer  than that for photospheric
lines. Using the last 6 profiles only improves the  fit slightly but
induces minimal changes in the resulting map.  We conclude that the 
recovered image is robust despite the low-level intrinsic variability.

The map of \caii\ excess emission at the surface of V2129~Oph shares
similarities  with the distribution of dark spots.  It shows in particular one
main feature close  to the pole and extending towards lower latitudes between
phase 0.80 and 0.95,  directly reflecting (i) the increased \caii\ emission at
these phases and (ii)  the fairly weak velocity change of the line emission
core.  Excess emission  from this polar accretion spot reaches about 1.5 times
that of the average quiet  chromosphere;  its relative area is about 5\% of the
total stellar surface.   Given the finite resolution of the imaging process,
the estimate we derive for the  fractional area covered by accretion spots is
likely an upper limit only.

\section{The large-scale magnetic topology}
\label{sec:zdi}

In this section, we use stellar surface imaging to model the Zeeman signatures 
from both LSD profiles and accretion proxies.  As in Sec.~\ref{sec:dim}, our 
aim is to demonstrate that the variability of Stokes $V$ signatures is due to 
rotational modulation and to obtain a consistent description of the large-scale 
field topology of V2129~Oph.  

\subsection{Model description}

The obvious difference between the sets of Zeeman signatures from the
photospheric LSD profiles  and the accretion proxies (see Sec.~\ref{sec:zee})
led us to suggest that photospheric and emission lines do not form over the
same regions of the stellar surface.   This is compatible with our findings of
Sec.~\ref{sec:dim} that the  accretion emission is concentrated in the dark
polar regions. The low surface brightness is expected to suppress the 
photospheric Zeeman signature, rendering the photospheric lines insensitive to
the field.   In the following, we assume this to be the case and  demonstrate
that both sets of Zeeman signatures are nonetheless consistent  with a
single large-scale magnetic topology.  

As in Sec.~\ref{sec:dim}, we use the \caii\ IRT lines as our preferred
accretion  proxy.  We describe it with the two-component model of
Sec.~\ref{sec:zee},  combining a basal chromospheric emission component evenly
distributed over the  star with an additional emission component concentrated
in local accretion  spots.   The emission-line Zeeman signatures are assumed to
be  associated with the accretion component only,  as supported by the
conclusions of Sec.~\ref{sec:zee}.  

The model we adopt involves a vector magnetic field $B$ and a  local accretion
filling-factor  $f$ which describes the local sensitivity to accretion  proxies
and photospheric lines\footnote{Note that the local accretion filling-factor  we
define here is different from the usual 'accretion filling factor' of the cTTS 
literature, i.e., the relative area of the total stellar surface covered by
accretion  spots.}. For $f=0$, the local area on the protostar's surface
contributes fully to photospheric  lines and generates no \caii\ excess
emission (and only unpolarised  chromospheric \caii\ emission);  for $f=1$, the
local area does not contribute at all to  photospheric lines and produces the
maximum amount of excess \caii\ emission and  polarisation (in addition to the
unpolarised chromospheric \caii\ emission).   Spectral contributions for
intermediate values of $f$ are derived through linear combinations between the
$f=0$ and $f=1$ cases.  We obtain both $B$ and $f$ by fitting the corresponding
synthetic Stokes $V$ profiles to the observed Zeeman signatures from both LSD
profiles and \caii\ emission cores. We can also incorporate the additional
constraint of fitting the unpolarised profiles of the \caii\ emission cores.  

The code we use for fitting $B$ and $f$ is adapted from the stellar surface
magnetic  imaging code of \citet{Donati01} and \citet{Donati06}.  In this
model, the  surface of the star is divided into a grid of thousands of small
surface pixels,  on which we reconstruct $B$ as a 3 component vector field and
$f$ as a simple  scalar field.  The magnetic vector field is described and
computed as a spherical-harmonic expansion, whose coefficients are determined
through fitting the simulated  Stokes $V$ profiles to the observed data sets.  
More specifically, the field is expressed as the sum of a poloidal field and a
toroidal  field, with the three field components in spherical coordinates
being  defined through three sets of complex spherical harmonics coefficients,
$\alpha_{\ell,m}$,  $\beta_{\ell,m}$ and $\gamma_{\ell,m}$. Here $\ell$ and $m$
denote the order and degree  of the spherical-harmonic mode\footnote{While
$\alpha_{\ell,m}$ describes the radial  field, $\beta_{\ell,m}$ describes the
non-radial poloidal field and $\gamma_{\ell,m}$  the toroidal field.   More
details about this description can be found in \citet{Donati06}.}.   

For a given magnetic topology, the unpolarised and polarised spectral
contributions  to photospheric lines are obtained through the Unno-Rachkovsky
model used in  Sec.~\ref{sec:dim}, and assuming a Land\'e factor and mean
wavelength of 1.2 and  620~nm respectively.  For the excess \caii\ emission
produced in the accretion region,  we assume the same Gaussian model as that of
Sec.~\ref{sec:dim}, with an  equivalent width set to twice that of the quiet
chromosphere profile (thus defining  the highest possible amount of local
excess emission), a unit  Land\'e factor and a mean wavelength of 850~nm.  

\begin{figure}
\center{\mbox{\includegraphics[scale=0.58,angle=-90]{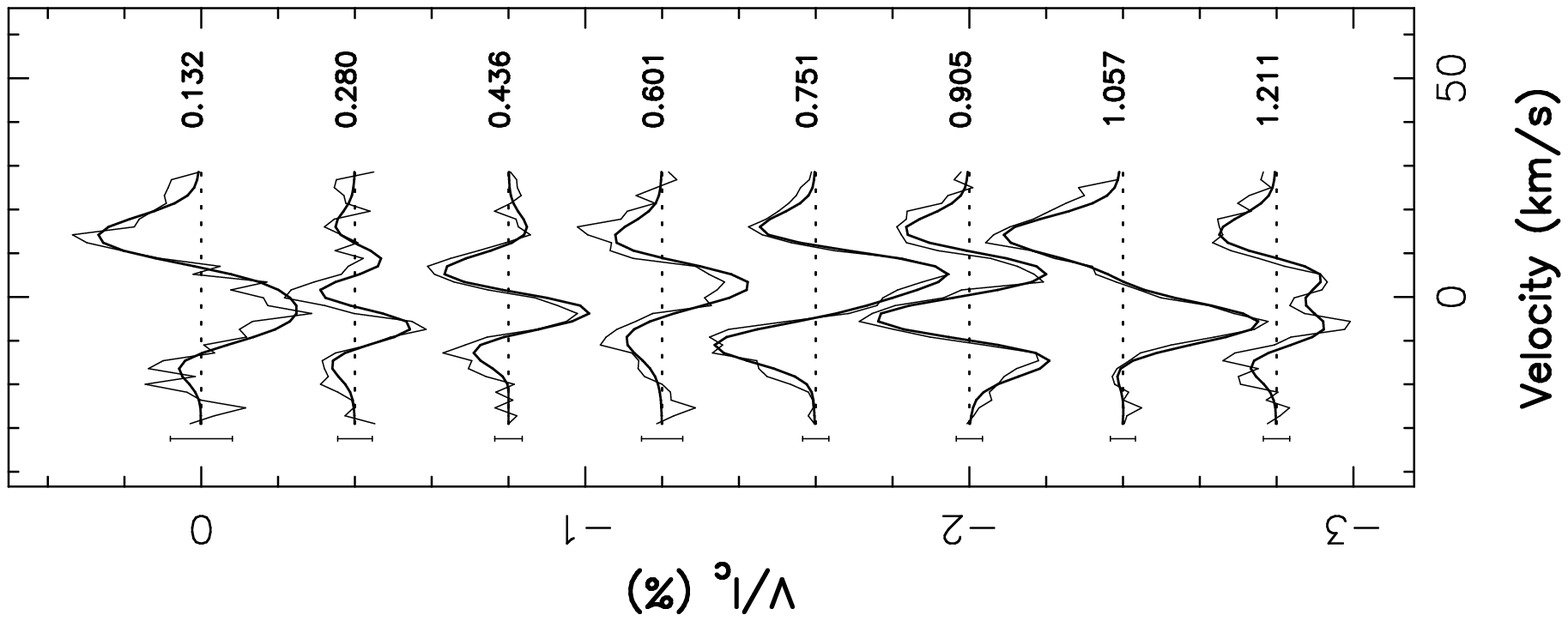}
              \includegraphics[scale=0.58,angle=-90]{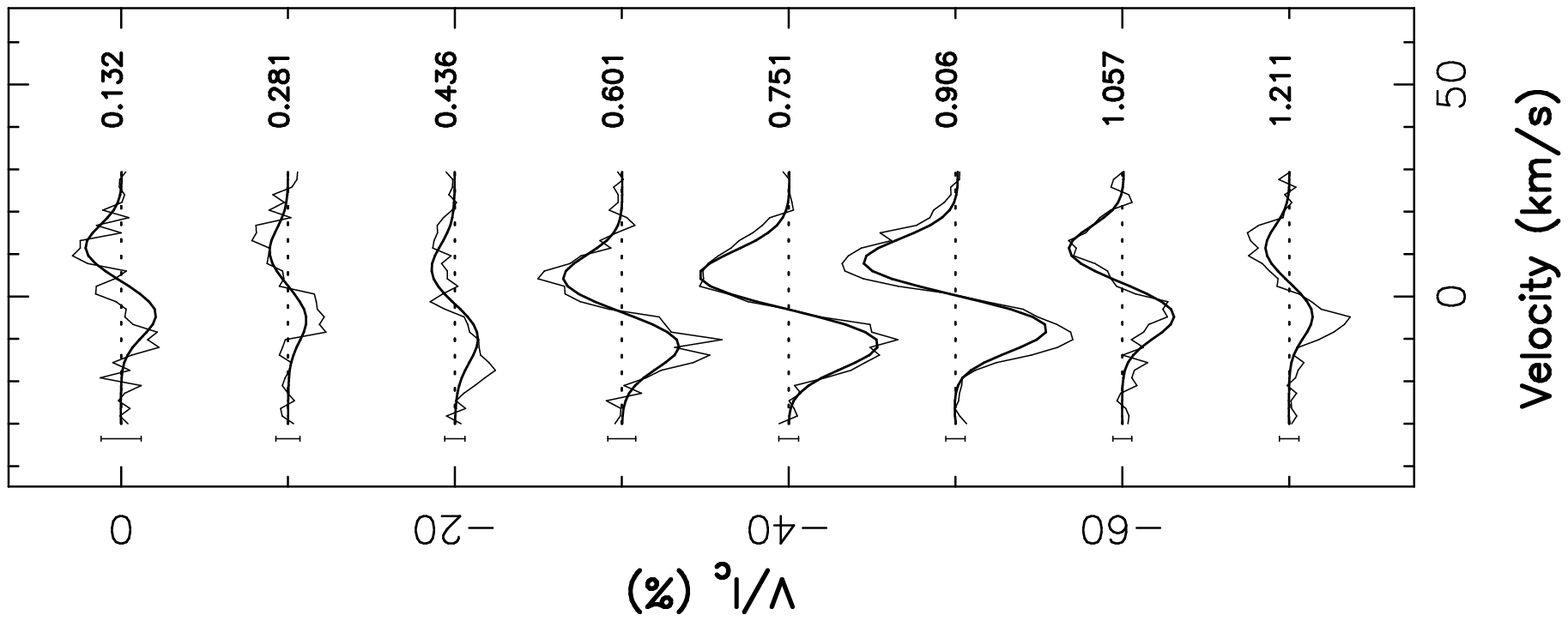}}} 
\caption[]{Stokes $V$ Zeeman signatures (thin line) from LSD profiles of
photospheric  lines (left column) and \caii\ emission cores (right column)
along with the  maximum entropy fit to the data (thick line).   The rotational
phase and cycle of each observation is written next to each profile.   A
3$\sigma$ error bar is also shown to the left of each profile.  }  
\label{fig:fitv}
\end{figure}

\begin{figure}
\includegraphics[scale=0.50]{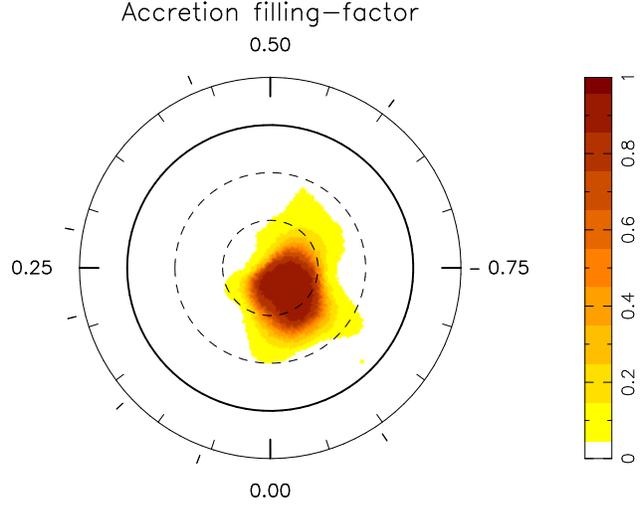}
\caption[]{Map of the local accretion filling factors at the surface of
V2129~Oph, as derived from  simultaneously fitting the two series of Zeeman
signatures (LSD profiles of  photospheric lines and \caii\ emission cores, see
Sec.~\ref{sec:zee}) and assuming  that each series senses a different spatial
region of the stellar surface (see  Sec.~\ref{sec:zdi}).  } 
\label{fig:acc}
\end{figure}

\begin{figure*}
\center{\mbox{\includegraphics[scale=0.88]{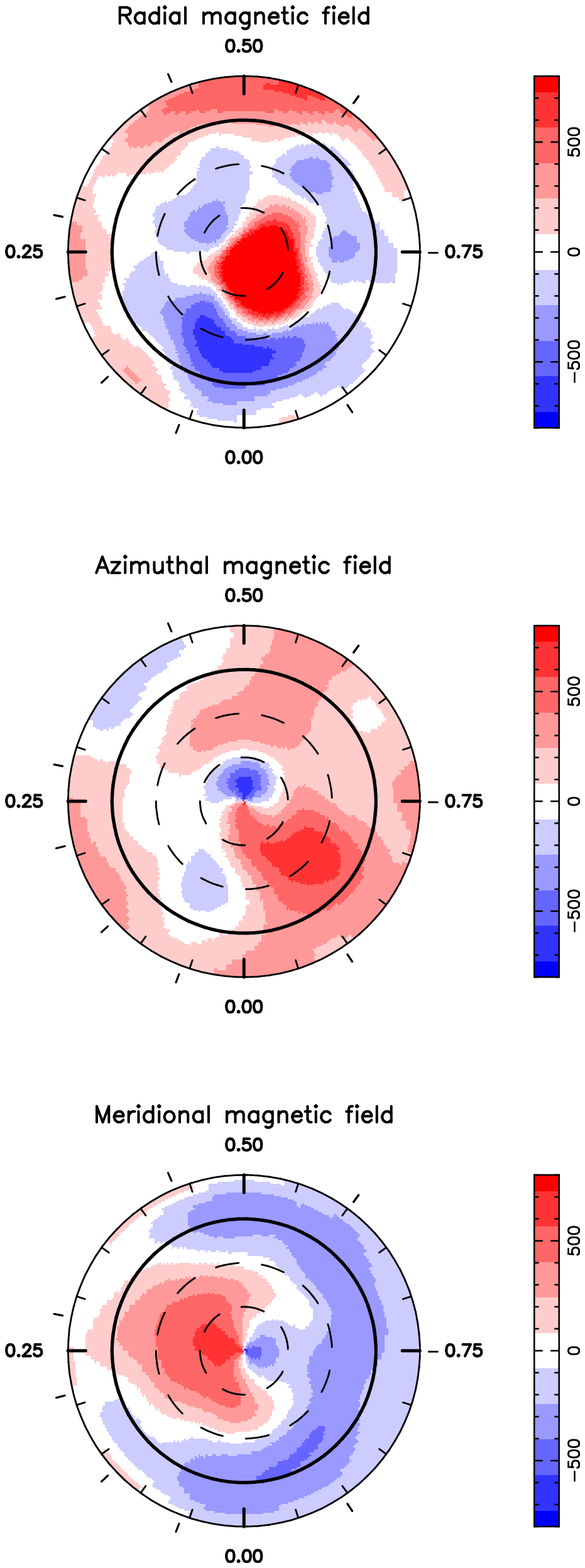}
              \includegraphics[scale=0.88]{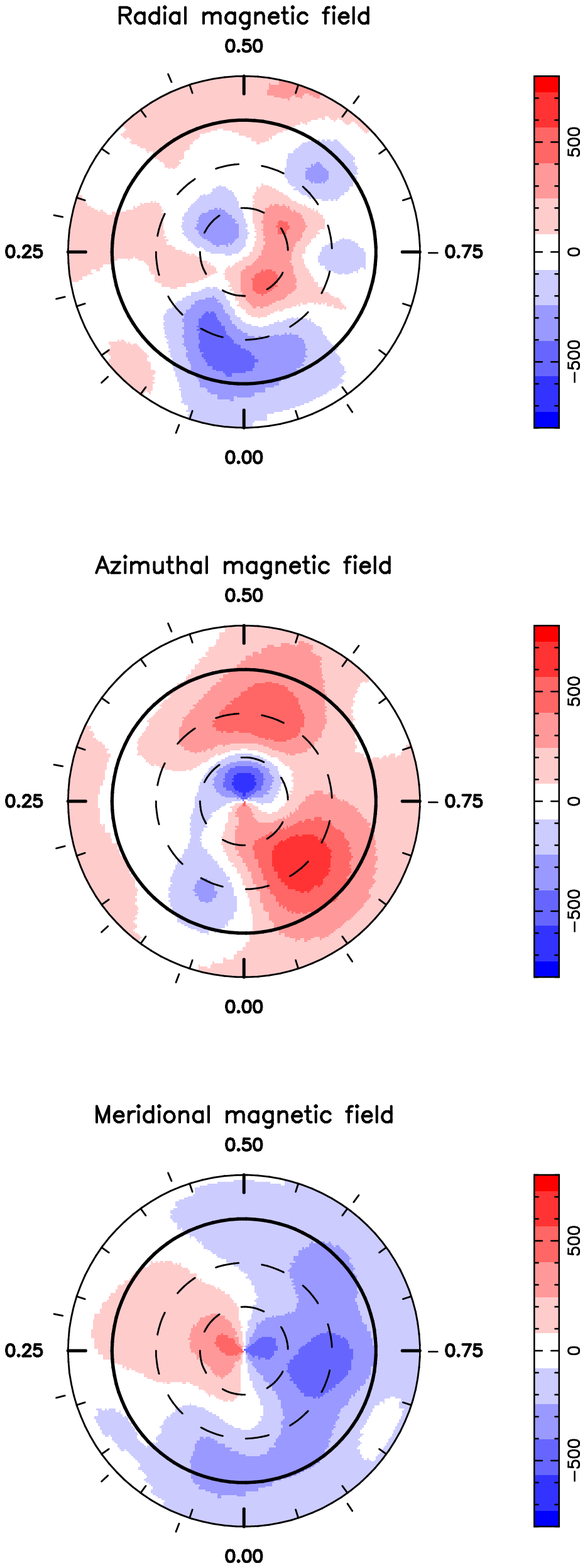}}} 	      
\caption[]{Reconstructed magnetic topology of V2129~Oph, simultaneously derived
from  our two series of Zeeman signatures (LSD profiles of photospheric lines
and \caii\  emission cores, see Sec.~\ref{sec:zee}, left panel) and when using
LSD Stokes $V$  profiles from photospheric lines only (right panel).  The image
on the right misses  most of the 2~kG high-latitude region of positive radial
field, in  which accretion occurs.  The three components of the field in
spherical  coordinates are displayed (from top to bottom) in both cases, with
field fluxes  labelled in G. } 
\label{fig:mapv}
\end{figure*}

\begin{figure*}
\center{\mbox{\includegraphics[scale=0.59]{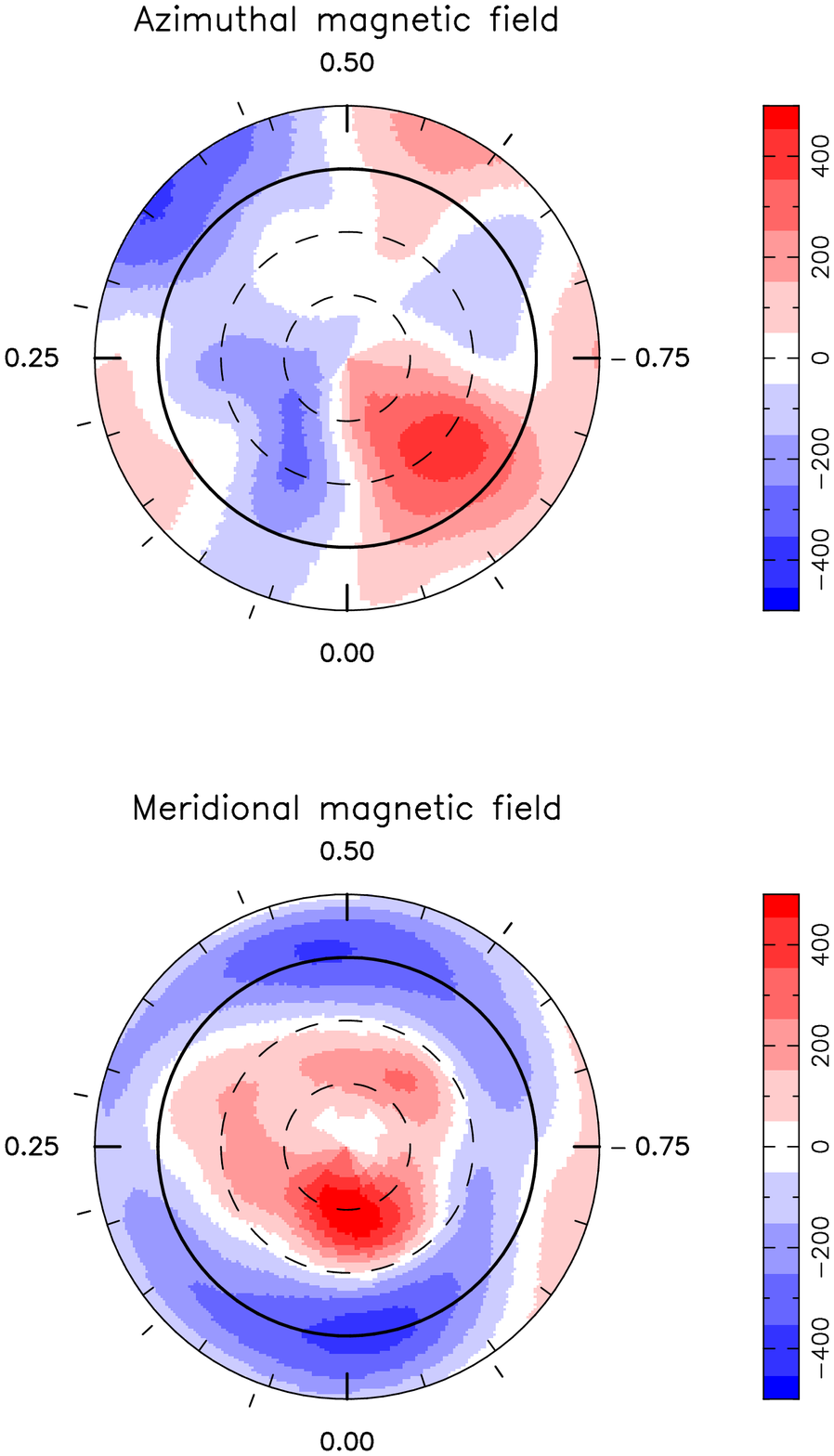}\hspace{2mm}
              \includegraphics[scale=0.59]{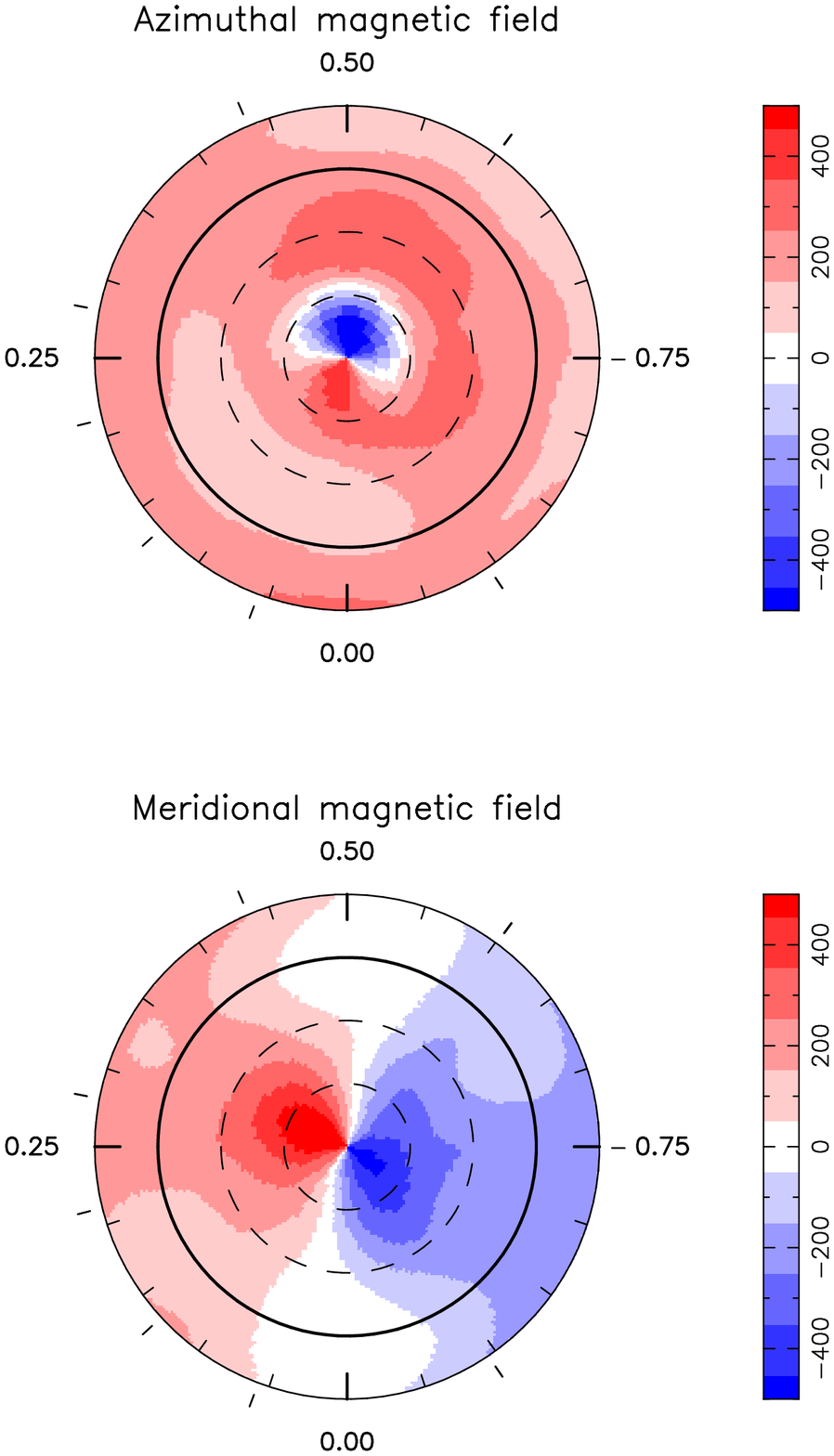}}}	      
\caption[]{Decomposition of the azimuthal and meridional components of the
reconstructed field (see Fig.~\ref{fig:mapv}, left column) into their
non-radial poloidal (left  column) and toroidal (right column) terms,
corresponding respectively to spherical-harmonic coefficients
$\beta_{\ell,m}$ and $\gamma_{\ell,m}$. } 
\label{fig:poltor}
\end{figure*}

Given the spatial resolution of the imaging process (see Sec.~\ref{sec:dim}),
we find it sufficient to limit the spherical-harmonic expansion of the magnetic
field components to terms  with $\ell\leq 20$ (230 modes altogether). In
practice, little improvement is obtained when adding spherical harmonic terms
corresponding  to $\ell\geq15$.  

\subsection{Modelling results}

We find that the Stokes $V$ signatures of both the photospheric LSD profiles
and the \caii\ emission cores  can be fitted simultaneously down to the noise
level using the model described  in Sec.~\ref{fig:fitv} above.  Repeating the
experiment for a range of rotation periods  (as we did in Sec.~\ref{sec:dim}
for Stokes $I$ profiles only) and assuming solid body  rotation yields optimal
fits to the Stokes $V$ data at a period of $6.62\pm0.07$~d,  again consistent
with the 6.53~d period derived by \citet{Shevchenko98}.  Assuming  now that the
star rotates differentially, we find that optimal fits to the Stokes  $V$
profiles are achieved for $\omeq=0.928\pm0.020$~\rpd\ and 
$\dom=-0.038\pm0.030$~\rpd, compatible within 3$\sigma$ with the values
derived  from the Stokes $I$ profiles.  This confirms our earlier conclusion
from inspection of the profiles in Sec.~\ref{sec:zee} that rotational
modulation dominates the temporal variability of  the Zeeman signatures.    

The maps of local accretion filling-factors and magnetic field derived from the
Stokes $V$ data are shown in  Figs.~\ref{fig:acc} and \ref{fig:mapv} (left
panel).  They confirm  and strengthen the conclusion that accretion is very
localised at the  stellar surface, being concentrated in a high-latitude region
where the radial field  reaches a peak value of 2.1~kG, in agreement with the
estimate derived in  Sec.~\ref{sec:zee}.  This accretion region occupies  about
5 percent of  the total stellar surface.  Including a simultaneous fit to the
unpolarised \caii\  emission profiles has little effect on the result, provided
we assume the  local \caii\ excess emission in the accretion regions to be
about twice the emission  from the quiet chromosphere.  Note that this
fractional area is only an upper limit.   A similar fit to the data can be
obtained by assuming that the local \caii\  excess emission in the accretion
spot is larger than that considered here. In this  case, we obtain a smaller
accretion hot spot whose average location is unchanged. The spatial resolution
is however high enough to claim that the accretion spot cannot  be larger than
the one we reconstruct.  

As a comparison, we also show  in Fig.~\ref{fig:mapv} (right panel) the
magnetic field topology reconstructed from  using the photospheric Stokes $V$
LSD profiles alone.   Unsurprisingly, this  topology is essentially the same as
that in the left-hand panel for all regions where no accretion occurs and for
which all available information comes from the photospheric lines.  This  map,
however,  misses most of the magnetic flux from the high-latitude accretion
spot, which now shows up as a weak positive field feature only.  

\begin{figure*}
\center{\hbox{\includegraphics[scale=0.34,angle=-90]{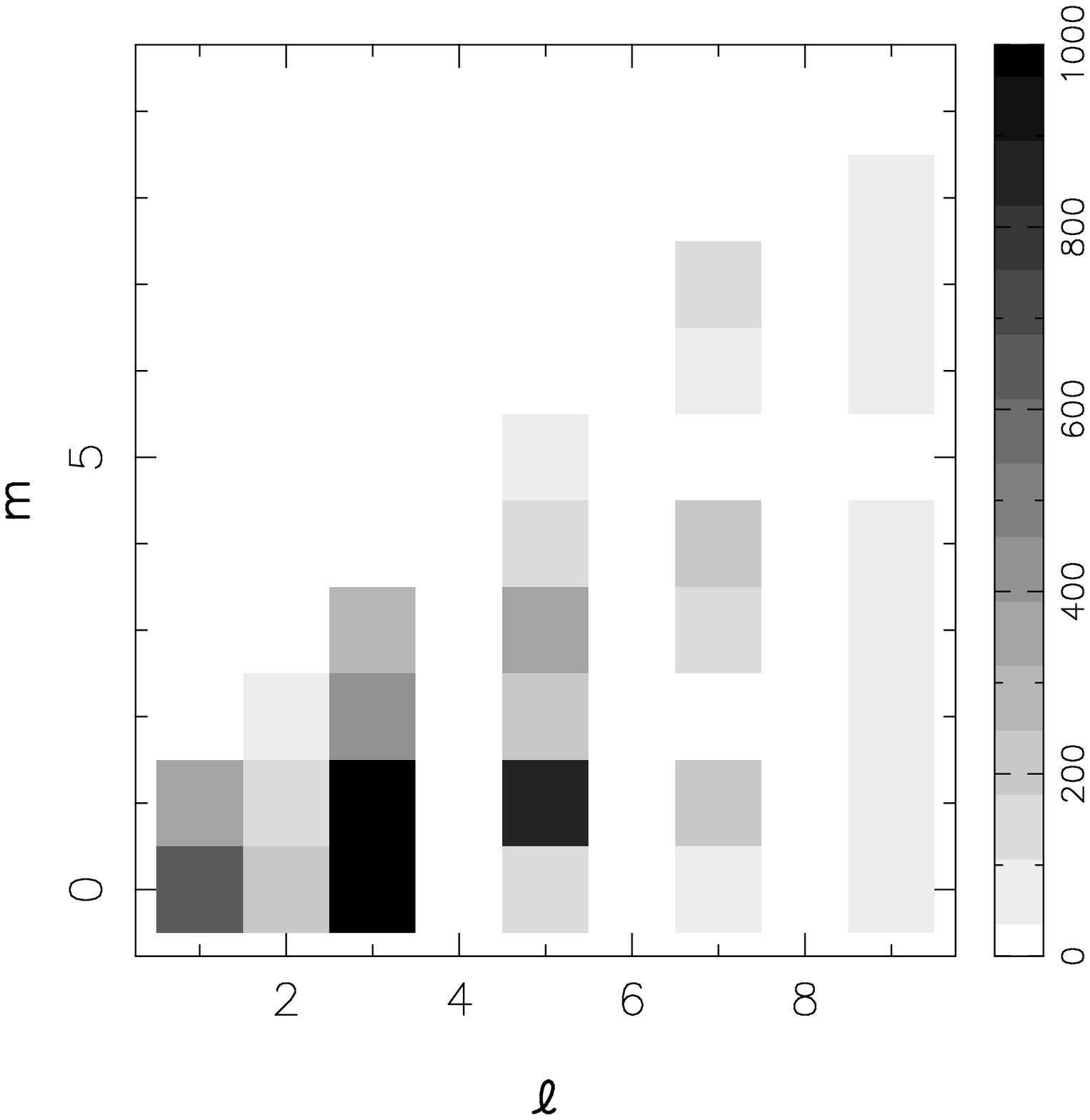}\hspace{2mm}
              \includegraphics[scale=0.34,angle=-90]{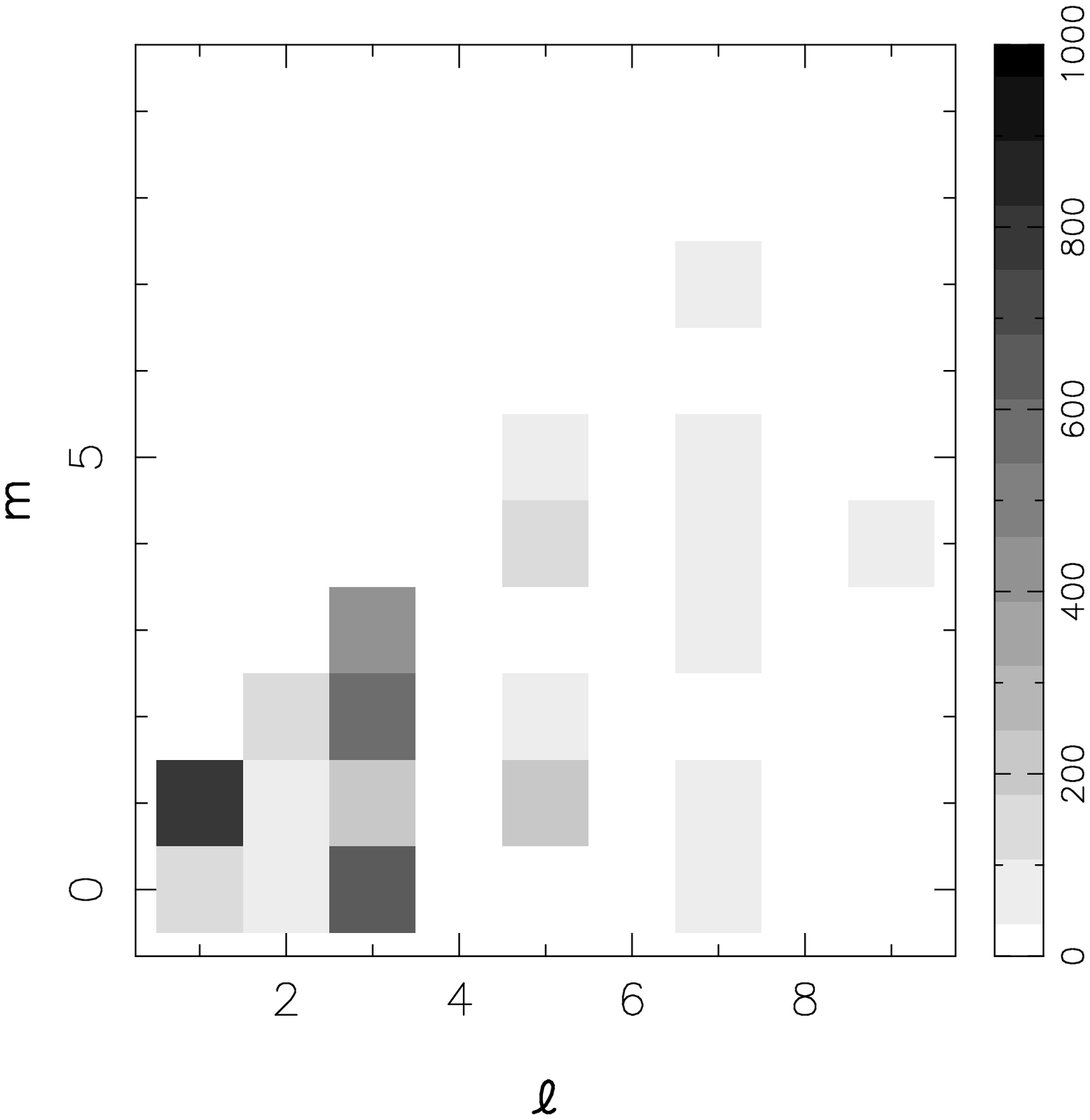}\hspace{2mm}
              \includegraphics[scale=0.34,angle=-90]{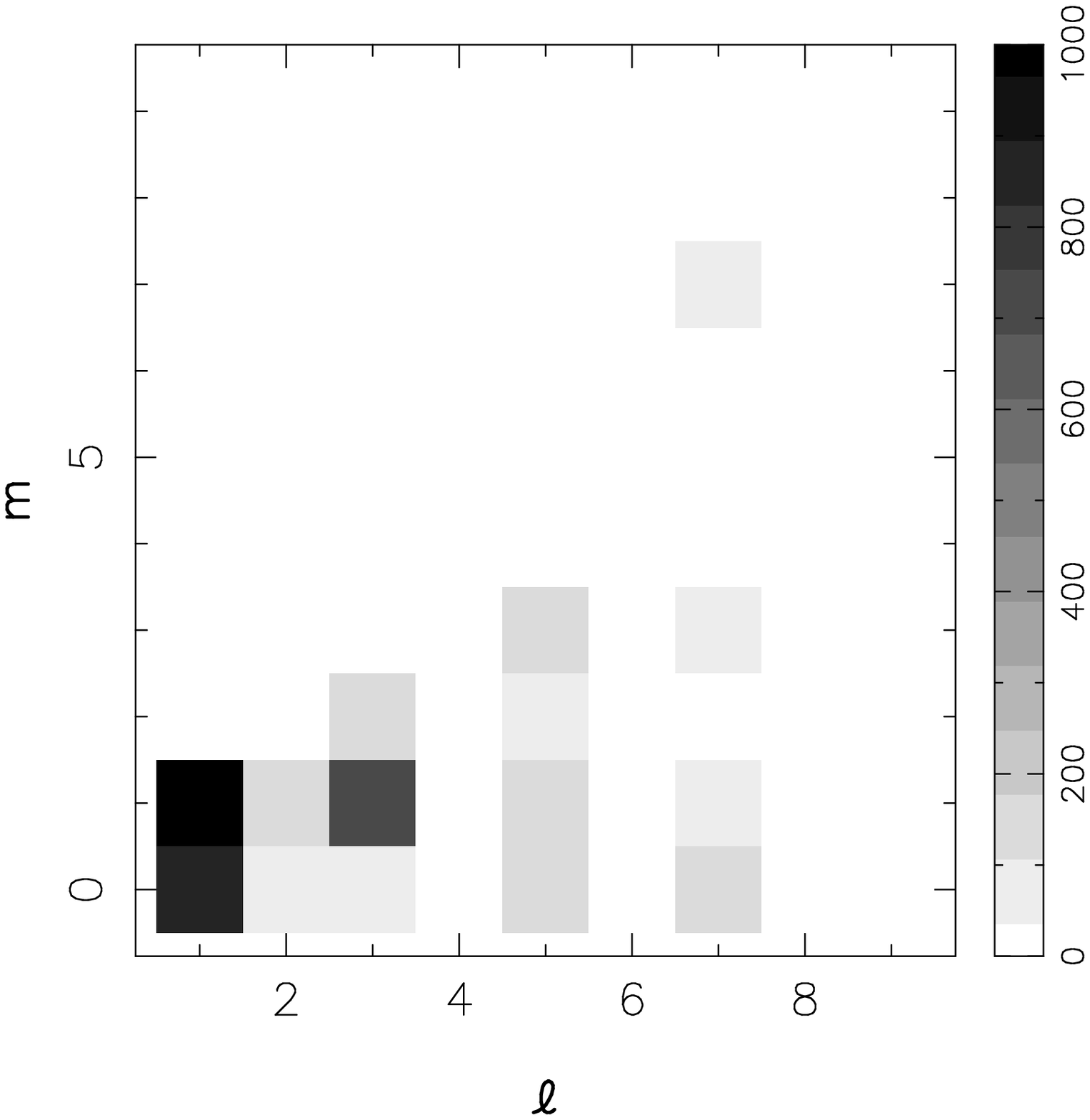}}}
\caption[]{Modulus (in G) of the spherical-harmonic complex coefficients 
$\alpha_{\ell,m}$, $\beta_{\ell,m}$ and $\gamma_{\ell,m}$ (from left to right, 
see text) for the reconstructed  field topology of V2129~Oph
(Fig.~\ref{fig:mapv}, left column),  as a function of mode degree $\ell$ and
order $m$.  In this case, the imaging code was  pushed towards antisymmetric
field configurations by favouring modes with odd $\ell$ values.   Modes with
$\ell<10$ only are displayed here. }
\label{fig:sph}
\end{figure*}

\subsection{Field topology}

Even though it is dominated by a 2~kG high-latitude positive radial field
feature,  the magnetic topology nevertheless differs significantly from a
dipole and exhibits  a number of clear specific features.   In particular, we
find that the fit to the data (and in particular to the Stokes $V$  LSD
profiles) is significantly poorer if we assume that the field is purely
poloidal,  implying that the toroidal field component we reconstruct is likely
to be real.   The spatial structure of this toroidal component is shown on
Fig.~\ref{fig:poltor}  (right column) along with the non-radial poloidal field
component (left column).   

The large-scale field topology includes a strong axisymmetric component.   The
poloidal field component features a strong positive radial field spot near the 
pole, surrounded by a negative radial field ring located slightly above the
equator  (see top left image of Fig.~\ref{fig:mapv}), suggesting a dominant
axisymmetric mode  with $\ell$ higher than 1. This is also visible from the
dual ring structure  in the meridional component of the poloidal field (see
bottom left image of  Fig.~\ref{fig:poltor}).  The toroidal field topology is
simple (see right panel of  Fig.~\ref{fig:poltor}) and comprises a ring of
counter-clockwise oriented field.  This unipolar ring is tilted with respect 
to the rotation axis towards phase 0.55 and passes through the rotation pole, 
giving it a complex multipolar appearance when plotted in spherical coordinates.  

Since the lower hemisphere of V2129~Oph is poorly constrained by observations 
(southern surface features being strongly limb-darkened and only visible for a
short  fraction of the rotation cycle), there is no direct way of working out
whether the  large-scale field topology is mainly symmetric or antisymmetric
with respect to the  centre of the star, or whether it features a mixed
combination of symmetric and antisymmetric modes.   By forcing the code towards
either symmetric or antisymmetric topologies or by letting  it choose by
itself, we find that all three options are possible and correspond to  magnetic
topologies that are very similar over the visible hemisphere and that fit the 
observations equally well.  While the mainly symmetric and antisymmetric
magnetic  topologies we reconstruct both include a strong field region near the
invisible pole  (mirroring that detected close to the visible pole, with equal
and opposite polarities  respectively), the third solution mixes both symmetric
and antisymmetric solutions in  roughly equal proportions and produces a
topology with no strong field region in the  invisible hemisphere.  Assuming
that magnetic topologies of cTTSs are either mainly  symmetric or antisymmetric
(most of them featuring strong unipolar field regions close  to the pole,
\citealt{Valenti04, Symington05}), we find that the mainly antisymmetric  field
topology is more likely as it requires a field with lower contrast for the
same quality of the fit to the data.  

In this antisymmetric case, the dominant mode of the reconstructed large-scale
field corresponds  to an octupole ($\ell=3$), with only little energy
reconstructed in the dipole field  configuration.  Fig.~\ref{fig:sph} shows how
the reconstructed magnetic energy  is distributed among the various modes. 
While the toroidal field term is mainly  confined to low order modes (mostly
$\ell=1$), the poloidal field extends to modes  with at least $\ell=5$ -- with
a maximum at $\ell=3$ -- and tends to concentrate mostly  in  axisymmetric
modes with small $m$ values.  Note that, given the limited 
resolution of our imaging process, we miss all magnetic flux at small spatial
scales  (e.g. in the form of closed bipolar groups) that may be present on
V2129~Oph.  

\begin{figure*}
\center{\hbox{\includegraphics[scale=0.42]{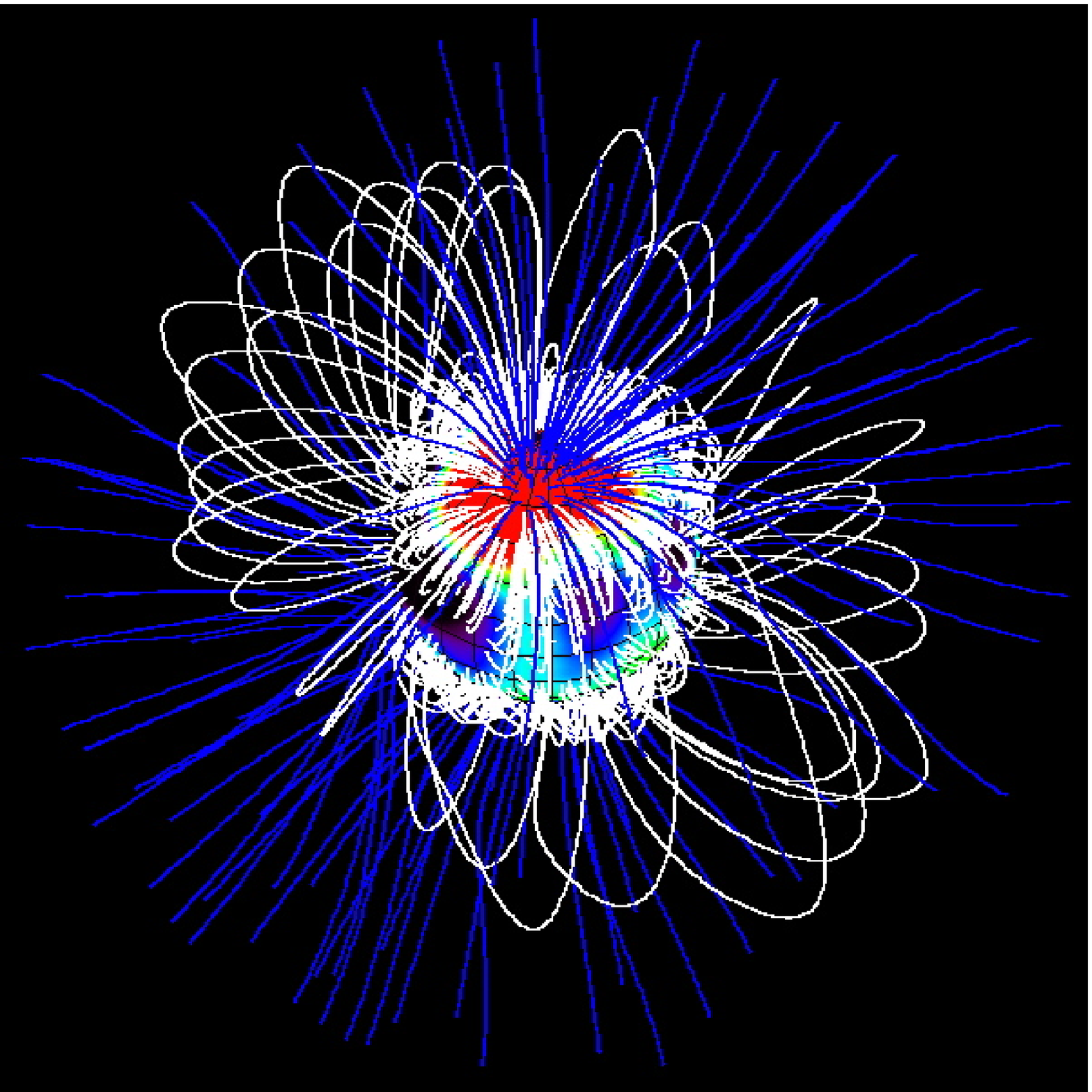}
              \includegraphics[scale=0.42]{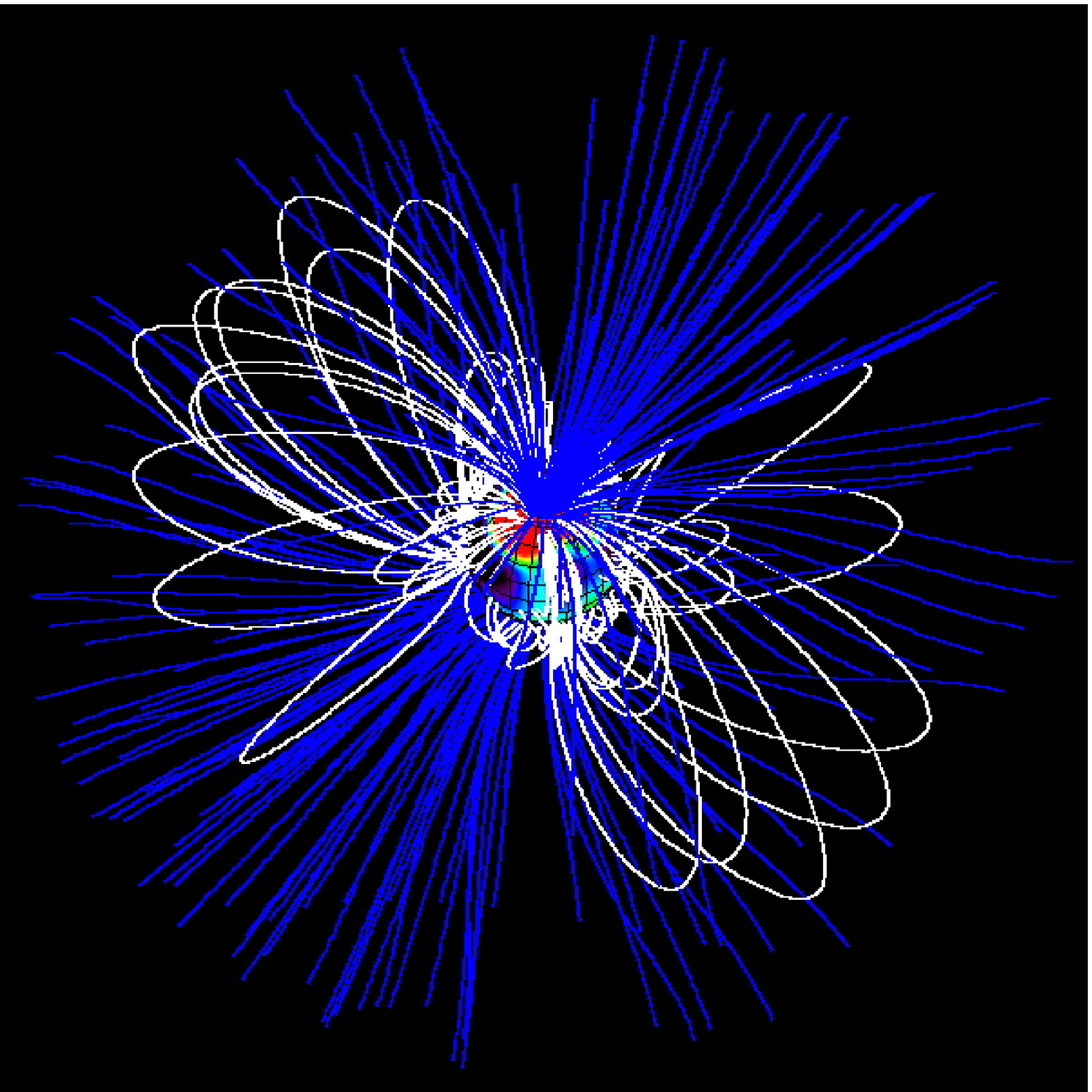}}}
\caption[]{Magnetospheric topology of V2129~Oph as derived from potential 
extrapolations of the reconstructed surface magnetic topology (left panel of 
Fig.~\ref{fig:mapv}).   The magnetosphere is assumed to extend up to the inner
disc radius, equal to  3.4 and 6.8~\rstar\ in the left and right panels
respectively.  The complex  magnetic topology close to the surface of the star
is very obvious.   In both cases, the star is shown at rotational phase 0.8. 
The colour patches  at the surface of the star represent the radial component
of the field  (with red and blue corresponding to positive and negative
polarities);  open  and closed field lines are shown in blue and white
respectively.  }
\label{fig:mag}
\end{figure*}

\begin{figure*}
\center{\hbox{\includegraphics[scale=0.42]{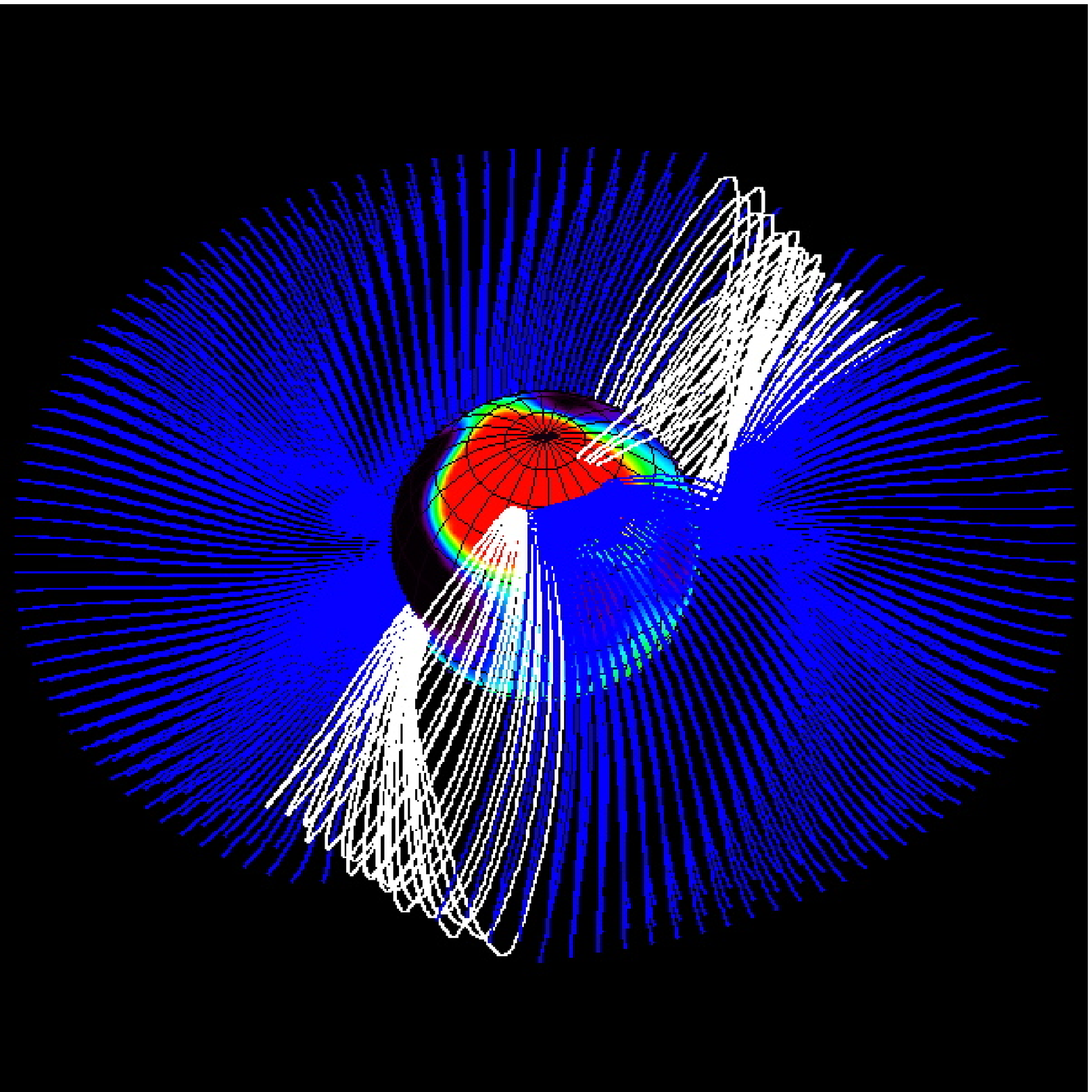}
              \includegraphics[scale=0.42]{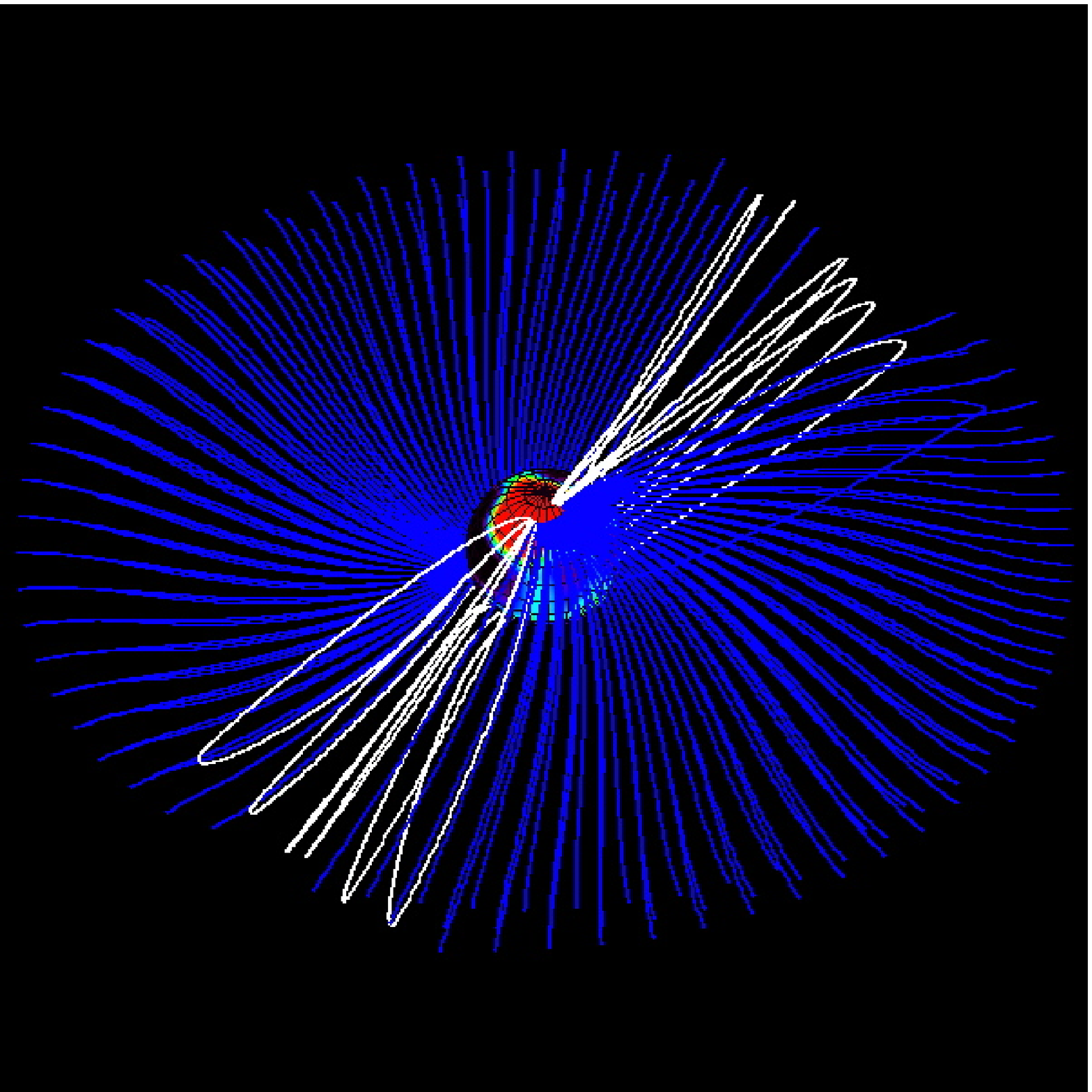}}}
\caption[]{Locations of field lines capable of accreting material from the
accretion  disc to the surface of V2129~Oph, as derived from the magnetospheric
topologies  shown in Fig.~\ref{fig:mag}.  Open field lines connecting to the
disc are shown in  blue, while closed field lines appear in white.  As in
Fig.~\ref{fig:mag}, the  magnetosphere is assumed to extend to either 3.4 or
6.8~\rstar\ (left and right  panels respectively).  The topology of the 
accretion columns becomes much more complex when the magnetosphere is assumed 
to be smaller. 
} 
\label{fig:cor}
\end{figure*}

\section{Magnetospheric accretion and corona}
\label{sec:mag}

To illustrate how accretion proceeds between the inner disc and the surface of
the  star, we extrapolate the reconstructed magnetic field over the whole
magnetosphere.   To do this, we assume that the 3D field topology is mainly
potential and becomes  radial beyond a certain magnetospheric radius \rmag\ from
the star, to mimic the opening of the largest magnetic loops under the coronal
pressure  \citep{Jardine02, Jardine06, Gregory06}.   In non-accreting stars,
this distance is usually assumed to be smaller than or equal  to the corotation
radius (\rcor) at which the Keplerian orbital period equals  the stellar
rotation period.  In cTTSs however, the magnetic field of the  protostar is
presumably clearing out the central part of the accretion disc,  suggesting
that the magnetosphere should extend as far as the inner disc  rim
\citep{Jardine06}.   The derived magnetospheric maps (see Fig.~\ref{fig:mag}
for two possible values of  \rmag) clearly show that the field topology is
complex, with an intricate network  of closed loops at the surface of the
star.  As \rmag\ is increased, the magnetosphere  is dominated by more extended
open and closed field lines.  

For several reasons, our description of the magnetosphere is only an
approximation.   The innermost field geometry is likely to be more complex than
our reconstruction, as small-scale magnetic structures at the surface of the
star remain undetected through  spectropolarimetry. This is not, however, a
crucial issue for the present study,  which focusses mainly on the medium- and
large-scale magnetic topology of V2129~Oph.   Moreover, the true magnetic
topology of V2129~Oph is likely not potential given  the strong plasma flows
linking the disc to the stellar surface and the significant  toroidal field
component detected at the surface of V2129~Oph.  Finally, the magnetic  field
in the disc is also expected to interact with the magnetosphere, at least in 
its outer regions \citep[e.g.,][]{vonRekowski04}.  These complications are not,
however, included in most existing theoretical studies on magnetospheric
accretion  \citep[e.g.,][]{Romanova03, Romanova04}.  We thus believe that,
despite its  limitations, our imaging study should be robust enough to yield
the rough locations of  accretion funnels and to provide a useful illustration
of the constraints  that spectropolarimetric data sets can yield.  More
quantitative analysis and  simulations are postponed to forthcoming papers.  

We can estimate where the accretion funnels  are located and where they are
anchored at  the surface of the star, by identifying those magnetospheric field
lines that are  able to accrete material from  the disc. Such field lines must
link the star to the disc and intersect the rotational equator  with effective
gravity pointing inwards in the co-rotating frame of reference 
\citep[e.g.][]{Gregory06, Gregory07}.  

If we assume that $\rmag=3.4$~\rstar, we find that disc material accretes
mainly on to the radial field spots located close to the equator,  with only a
small fraction of the accretion flow reaching the high-latitude spots. This is
obviously in strong contradiction with the conclusions of Sec.~\ref{sec:zdi}.  
If we assume a more extended magnetosphere with $\rmag=6.8$~\rstar, most of
the accreting material flows onto the high-latitude accretion spots.  Pushing
\rmag\ further out (to 7.5~\rstar) focusses  all the accreting material at high
latitudes only, in much better agreement with  observations (see
Figs~\ref{fig:cor} and \ref{fig:ftp}).   We therefore conclude that \rmag\ is
closer to 7~\rstar\ in V2129~Oph  and likely coincides with \rcor, equal to 
$0.075\pm0.003$~AU or $6.8\pm1.2$~\rstar.  

In the magnetospheric topology we derive, the northern accretion funnel passes 
in front of the high-latitude accretion hot spot between rotation phases 0.7 and 
0.8.  We suggest that free-falling material in this accretion funnel is responsible 
for the high-velocity absorption component observed in the red wing of Balmer 
lines at cycle 0.75 (see Sec.~\ref{sec:zee}), just as solar prominences transiting 
the solar disc are 
viewed as dark filaments.  Similarly, we propose that free-falling plasma along 
the southern accretion funnel that plunges into the invisible accretion spot 
produces, through scattering, the redshifted emission component that Balmer lines 
feature at cycle 1.21.  The orientation of the southern accretion funnel is indeed 
ideally oriented for this purpose at this particular rotation phase.  Detailed 
radiative transfer computations \citep[e.g.,][]{Symington05b, Kurosawa06} 
are needed to confirm this.

The magnetospheric field geometry derived from the reconstructed field maps can
also be used to model the corona of V2129~Oph and  estimate the associated
X-ray flux.  Assuming $\rmag=7$~\rstar\ and an isothermal corona  at a
temperature of 20~MK (typical for the hot component of X-ray emission from 
cTTSs) filled with plasma in hydrostatic equilibrium \citep[e.g.][]{Jardine06, 
Gregory06}, we obtain an X-ray emission measure of $4\times10^{53}$~\pcc\ and
an average (i.e., emission-measure-weighted) coronal density of
$5\times10^{9}$~\pcc.   This value is typical for solar-mass cTTSs such as
V2129~Oph according to the statistics  derived from the COUP (Chandra Orion
Ultradeep Project) survey \citep[e.g.,][]{Jardine06}.    It translates into a
X-ray luminosity of $4\times10^{30}$~\eps, matching well the published  value
of $2.5\times10^{30}$~\eps\ \citep{Casanova95}.

\section{Discussion}
\label{sec:dis}

The spectropolarimetric data we have collected on V2129~Oph yields the  first
realistic model of the large-scale magnetic topology on a mildly accreting 
cTTS.  Although this first study of its kind addresses only one  star,
V2129~Oph can nevertheless be  regarded as prototypical of mildly accreting
cTTSs,  exhibiting no behaviour that would mark this star out as  peculiar.  We
can thus use it to readdress a number of issues regarding cTTSs.  

\subsection{Cool spots and hot accretion regions}

We observe that profile variations due to intrinsic variability (e.g.,
resulting from  unsteady accretion flows) are smaller than those caused by
rotational modulation, for  both the photospheric lines and the  accretion
proxies.  This conclusion is already  robust and confirms previous published
results based on data sets with similar time  sampling and coverage to ours
\citep[e.g.,][]{Valenti04}.  Future observations spanning timescales of at
least 2 complete rotation cycles, would be welcome to  establish whether this
situation is typical.  

The photospheric line profiles of V2129~Oph are variable in shape and to a
lesser extent in  strength.   The profile distortions consist of bumps (rather
than dips) and can be attributed  to the presence of cool spots at the surface
of the star, rotating in and out of the  observer's view.   This confirms the
conclusions of \citet{Shevchenko98}, who deduced from the photometric  colour
variations  that the brightness inhomogeneities on V2129~Oph are mostly  due to
cool and dark spots rather than to bright and hot regions.   With Doppler
imaging,  we successfully reproduce the rotational modulation of photospheric
line shapes and  find that the surface spot distribution is dominated by a cool
region over the visible  pole.  In this respect, this spot distribution differs
from those of cool active stars  with similar characteristics, which usually do
not show polar features at  rotation periods longer than about one week (e.g.,
the K1 evolved RS~CVn subgiant II~Peg,   \citealt{Petit06}).   

Accretion spots are also present at the surface of the star.  Their spectral
signatures are not seen at photospheric level, but  show up clearly in emission
proxies such as the \caii\ IRT and \hei\ lines which usually  trace heating of
the star's upper atmospheric layers. 

These emission proxies  include two components: a stationary emission component
presumed to originate in a  quiet chromosphere covering the whole stellar
surface, and a rotationally-modulated  emission feature presumably formed in
local accretion spots.  The accretion component  exhibits only small-amplitude 
radial velocity modulation, indicating that accretion spots are  located at
high latitudes.  This is confirmed through tomographic imaging, which shows
that accretion on the visible hemisphere of V2129~Oph is concentrated in a
single  high-latitude spot covering less than 5\% of the stellar surface,
located within  the main cool polar spot detected at photospheric level.  It
suggests that the heat  produced in the accretion shock is not transferred to
the photosphere efficiently  enough to warm it up above the temperature of the
non-accreting photosphere.  We  speculate that a second accretion spot is
present near the invisible pole, at  a location symmetrical  with respect to
the centre of the star.  

We also find that differential rotation of V2129~Oph is comparable with or less
than  that of the Sun, in agreement with previous observations \citep{Johns96}
and theoretical predictions \citep{Kuker97}.  Our result invalidates in
particular early suggestions that  cTTSs might be strong differential rotators
\citep{Smith94}.  

\begin{figure}
\center{\includegraphics[scale=0.48]{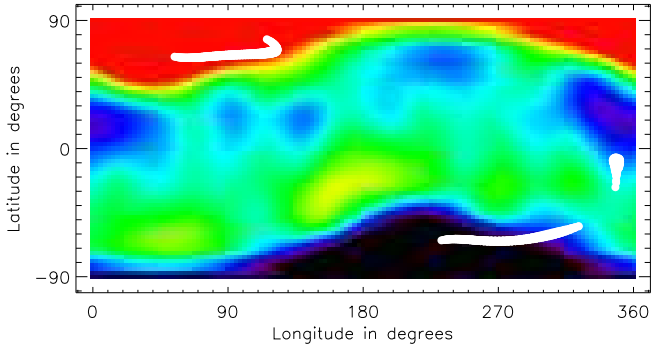}} 
\caption[]{Location of the footpoints of accreting field lines at the surface
of the  star (white circles), assuming that the magnetosphere extends to
6.8~\rstar.  In this  case, accreting field lines concentrate almost
exclusively in the strong unipolar radial field regions at high latitudes. 
The colours at the surface of the star depict  the radial field component (with
red and blue corresponding to positive and negative  polarities).  Note that
phase runs backwards with longitude, i.e., decreases from 1 to 0  while
longitude increases from 0\degr\ to 360\degr.  }
\label{fig:ftp}
\end{figure}

\subsection{Magnetic topology and origin of the field}

We observe that rotational modulation dominates the temporal evolution of the
Stokes  $V$ profiles of both photospheric lines and accretion proxies. 
Comparing Zeeman  signatures from both data sets suggests that they form in
different complementary regions of the stellar surface, with accretion proxies
tracing magnetic fields  at the footpoints of accretion funnels and
photospheric lines diagnosing fields in  the non-accreting photosphere.  With
this assumption, we are able to reconstruct,  through tomographic imaging, a
consistent magnetic topology that simultaneously  reproduces both sets of
Zeeman signatures.  

Despite being anchored in 2~kG high-latitude radial field spots, the
reconstructed  magnetic field is more complex than the simple dipole assumed in
most theoretical  models of magnetospheric accretion
\citep[e.g.,][]{Romanova03, Romanova04}.  The  negative radial field ring
encircling the main positive radial field pole provides  direct evidence that
the large-scale poloidal field structure corresponds to  spherical harmonics
order $\ell\geq2$.  We find that the field of V2129~Oph  is more likely to be
mainly antisymmetric (rather than symmetric) with respect to  the centre of the
star;  the respective contributions to the 2~kG radial field region  from the
$\ell=1$ (tilted dipole), $\ell=3$ (tilted octupole) and $\ell=5$ terms are 
about 0.35, 1.2 and 0.6~kG.  The reconstructed field also includes a toroidal
field  component in the form of a tilted ring of counterclockwise azimuthal
field.  

Given that fossil magnetic fields are presumably dissipated very quickly in
largely  convective stars \citep[e.g.,][]{Chabrier06}, the most natural
assumption is that  large-scale fields of cTTSs are generated through dynamo
action.  The mainly  axisymmetric field reconstructed on V2129~Oph argues in
this direction.  The magnetic topology seen here, however, differs in 
important respects from those of moderately rotating cool  active stars with
similar characteristics \citep[e.g.,][]{Petit06}.  The strong  high-latitude
unipolar field region detected on V2129~Oph and the relatively strong 
$\ell\geq3$ orders of the large-scale field are the most obvious.  These 
differences suggest that cTTSs may feature a composite field, including both  
fossil and dynamo components.  More data are needed to confirm this and to
explore  the potential impact on stellar formation.  

\subsection{Disc-star magnetic coupling}

Several theoretical papers \citep[e.g.,][]{Konigl91, Shu94, Cameron93} studied 
how the stellar magnetic field interacts with the surrounding accretion disc
and disrupts its vertical structure close to the star. They showed further that
the balance between accretion torques and angular momentum losses causes the
rotation of the star to evolve towards an equilibrium in which the disc
disruption  radius lies close to \rmag\ and just inside the co-rotation radius 
\rcor.  Assuming different models of the star/disc interaction, these authors 
derived convenient expressions  for \rmag;  they proposed that this coupling 
causes cTTSs to slow down to the  Keplerian orbital period at \rmag\ (thus 
equating \rmag\ and \rcor), explaining  why cTTSs are on average rotating more 
slowly than their disc-less equivalents.  This scenario is refered to as 
'disc-locking' in the literature.  

These  papers can be used to estimate \rmag\ as a function of the mass, radius,  
mass-accretion rate and magnetic dipole strength of the protostar.  Using values 
listed in Sec.~\ref{sec:par} and a dipole strength of 0.35~kG (see Sec.~\ref{sec:zdi}), 
we find that \rmag\ can be as low as 2~\rstar\ if using the model of \citet{Konigl91}  
or \citet{Shu94}, i.e., far smaller than \rcor\ (about 7~\rstar).  The model of 
\citet{Cameron93}, yielding \rmag\ values of  5.1 to 8.1~\rstar\ (in better 
agreement with \rcor), appears to provide a better  quantitative description of the 
star/disc magnetic interaction assuming this  coupling is responsible for the slow 
rotation of V2129~Oph.  

The disc-locking scenario was recently criticised on various grounds.  
Observations revealed that surface magnetic strengths on cTTSs (measured from Zeeman 
magnetic broadening of unpolarised profiles) do not correlate well with those 
expected from theoretical models, casting doubts on the validity of the 
disc-locking mechanism \citep[e.g.,][]{Bouvier07, Johns07}.  On theoretical grounds, 
authors claimed that a wind from the protostar would blow the field open at 3~\rstar\ 
\citep[e.g.,][]{Safier98, Matt04} and prevent magnetic coupling between the star 
and disc.  

Our study however demonstrates that the most relevant physical quantity for 
studying this issue is not the surface magnetic field itself, but rather the 
large-scale dipole strength on which magnetic coupling mostly depends.  On 
V2129~Oph at least, the large-scale dipole component is compatible with the 
predictions of \citet{Cameron93}.  Moreover, our observations establish that 
accretion occurs onto the polar regions of V2129~Oph rather than the equator, 
a phenomenon which we can only explain if the high-latitude field links to 
the accretion disc at a distance of about 7~\rstar.  This is obviously not 
compatible with the field being blown open by the wind at 3~\rstar.

\section{Conclusion}

In this paper, we report the discovery of magnetic fields on the cTTS 
V2129~Oph using ESPaDOnS, the new stellar high-resolution spectropolarimeter 
recently installed at CFHT.  Circular polarisation Zeeman signatures are
detected  in photospheric lines and in emission lines tracing magnetospheric
accretion.   We demonstrate that the temporal variations of both unpolarised
and circularly  polarised profiles, monitored over the complete rotational
cycle of V2129~Oph,  are mostly attributable to rotational modulation.  

From our sets of Zeeman signatures from photospheric lines and accretion 
proxies simultaneously, we recover the medium- and large-scale magnetic
topology  at the surface of V2129~Oph using tomographic imaging tools.  In this
process, we  also successfully derive the location of the accretion regions,
tracing the  footpoints of accretion funnels at the surface of the star.   We
find that the magnetic topology of V2129~Oph is significantly more complex 
than a dipole and is dominated by a 1.2~kG octupole, tilted by about  20\degr\
with respect to the rotation axis.  The large-scale dipole component  is much
smaller, with a polar strengh of only 0.35~kG.  The magnetic field  is mainly
antisymmetric with respect to the centre of the star, and the accretion 
regions are found to coincide with the two main high-latitude magnetic poles 
(each covering about 5\% of the total stellar surface).  The high-latitude 
accretion spots apparently coincide with dark polar features at  photospheric
level.  The magnetic topology of V2129~Oph is unusually complex  compared to
those of non-accreting cool active subgiants with moderate rotation  periods.  

As an illustration, we also provide a first attempt at modelling the
magnetospheric  topology and accretion funnels of V2129~Oph using field
extrapolation.   We find that the magnetosphere must extend to distances of
about 7~\rstar\ to  produce accretion funnels that match the observations and
anchor in the high-latitude  accretion spots identified at the stellar
surface.  This distance is roughly equal  to the corotation radius and matches
the theoretical predictions of \citet{Cameron93},  suggesting that the
star/disc magnetic coupling can possibly explain the slow rotation  of
V2129~Oph.  It suggests that magnetic field lines from the protostar are
apparently  capable of coupling to the accretion disc beyond 3~\rstar.  The
accretion geometry  we derive is qualitatively consistent with the modulation
of Balmer lines and  consistent with the X-ray coronal fluxes typical of 
cTTSs.  

More similar data sets are required to check how the conclusions of this
prototypical  study apply to other cTTSs, with different masses, rotation and
accretion rates.

\section*{Acknowledgements}

We thank the CFHT staff for their help during the various runs with ESPaDOnS.


\bibliography{v2129oph}

\bibliographystyle{mn2e}

\end{document}